\documentclass[12pt, draftclsnofoot, onecolumn]{IEEEtran}

\usepackage{amsmath}
\usepackage{amssymb}
\usepackage{graphicx}
\usepackage{float}
\usepackage{fixltx2e}
\usepackage{subeqnarray}
\usepackage{cases}
\allowdisplaybreaks
\IEEEoverridecommandlockouts
\usepackage[square, comma, sort&compress, numbers]{natbib}
\usepackage{multirow}
\usepackage{stfloats}
\usepackage{fixltx2e}
\usepackage[font={small}]{caption}
\usepackage{epsfig}
\usepackage{epstopdf}

\newcommand{\Tr}{\mathrm{Tr}}

\newtheorem{Theorem}{Theorem}

\ifCLASSINFOpdf
\else
\fi
\usepackage{amsmath}
\usepackage{array}
\usepackage{fixltx2e}

\usepackage{stfloats}

\newtheorem{theorem}{Theorem}

\newtheorem{corollary}[theorem]{Corollary}

\begin{document}
%
\title{An Analysis of Uplink Asynchronous Non-Orthogonal
Multiple Access Systems}

\author{\IEEEauthorblockN{Xun Zou, \textit{Student Member, IEEE}, Biao He, \textit{Member, IEEE}, and Hamid~Jafarkhani, \textit{Fellow, IEEE}}\\
	\thanks{Results in this paper were presented in part at the IEEE International Conference on Communications (ICC) 2018 \cite{xunICC18}. This work was supported in part by the NSF Award CCF-1526780.}
		\thanks{The authors are with the Center for Pervasive Communications and Computing, Department of Electrical Engineering and Computer Science, University of California, Irvine, CA, 92697 USA (email: \{xzou4, biao.he, hamidj\}@uci.edu).}
}

\maketitle

\begin{abstract}
	Recent studies have numerically demonstrated the possible advantages of the asynchronous non-orthogonal multiple access (ANOMA) over the conventional synchronous non-orthogonal multiple access (NOMA). The ANOMA makes use of the oversampling technique by intentionally introducing a timing mismatch between symbols of different users. Focusing on a two-user uplink system, for the first time, we analytically prove that the ANOMA with a sufficiently large frame length can always outperform the NOMA in terms of the sum throughput. To this end, we derive the expression for the sum throughput of the ANOMA as a function of signal-to-noise ratio (SNR), frame length, and normalized timing mismatch. Based on the derived expression, we find that users should transmit at full powers to maximize the sum throughput. In addition, we obtain the optimal timing mismatch as the frame length goes to infinity. Moreover, we comprehensively study the impact of timing error on the ANOMA throughput performance. Two types of timing error, i.e., the synchronization timing error and the coordination timing error, are considered. We derive the throughput loss incurred by both types of timing error and find that the synchronization timing error has a greater impact on the throughput performance compared to the coordination timing error.
\end{abstract}

\begin{IEEEkeywords}
Non-orthogonal multiple access, asynchronous transmission, oversampling, timing mismatch, interference cancellation.
\end{IEEEkeywords}

\section{Introduction}\label{sectionIntro}

Non-orthogonal multiple access (NOMA) is envisaged as a promising technique for future radio access \cite{ding2017survey}. Traditional orthogonal multiple access (OMA) techniques allocate orthogonal resources to different users, e.g., orthogonal time resources in the time division multiple access (TDMA) scheme. Differently, the NOMA provides the multiuser access by allocating non-orthogonal resources to users~\cite{liu2017downlink}. For example, in the power-domain NOMA scheme, the signals for multiple users are superposed at different power levels using superposition coding \cite{tse2005fundamentals}, and the multiuser detection method, such as successive interference cancellation (SIC) \cite{jafarkhani2005space}, is employed at the receiver. The advantages of the NOMA over the OMA have been extensively studied in~\cite{ding2017survey} and the references therein, e.g., providing higher system throughput compared to OMA and supporting massive connectivity.

Another line of research is to study the effects of asynchronous transmission on the performance of the wireless communication systems. Asynchronous transmission refers to the case where the symbol epochs of the signals transmitted by the users are not aligned at the receiver~\cite{verdu1989capacity}. In particular, \cite{verdu1989capacity} first pointed out the potential advantages of symbol-asynchronous communications in terms of increasing the capacity of a multiple-access channel. The work in \cite{xunGlobecom16} applied the symbol-asynchronous channel estimation method to tackle the pilot contamination problem in massive multiple-input multiple-output (MIMO) systems. Asynchronous transmission was studied in \cite{ganji2016interference,haci2017performance} as a tool to mitigate or cancel the inter-user interference. In addition, the nonzero symbol offset was used to reduce the inter-antenna interference in MIMO systems in~\cite{das2011mimo}. Moreover, an asynchronous analog network coding scheme for multiuser cooperative communications was proposed in \cite{zhang2017exploiting} to provide a greater diversity order compared with that of synchronous analog network coding. Also, adding intentional timing mismatch was proposed in \cite{zhang2017asynchronous} to improve the performance of a relay network. The authors of \cite{poorkasmaei2015asynchronous,avendi2015differential} further proposed several differential decoding schemes for asynchronous multiuser MIMO systems based on orthogonal space-time block codes (OSTBCs) and for differential distributed space-time coding systems with imperfect synchronization.

Applying the symbol-asynchronous transmission to NOMA, a scheme named asynchronous NOMA (ANOMA) was studied in \cite{cui2017asynchronous}. In fact, an idea similar to the ANOMA in \cite{cui2017asynchronous}, i.e., applying asynchronous transmission for multiple access, has also been proposed and investigated in, e.g., \cite{zhang2017asynchronous,ganji2016interference,haci2017performance}. Specially, a timing mismatch between signals for different users is intentionally added as an additional resource to address the problem of inter-user interference. It has been shown using the numerical simulation in \cite{cui2017asynchronous} that the ANOMA outperforms the conventional (synchronized) NOMA by achieving a larger throughput. 

However, the work in~\cite{cui2017asynchronous} has several limitations. While addressing those limitations is important to understand the ANOMA systems, to the best of our knowledge, no existing paper has tackled the following issues. 
First, there is no analytical result on the comparison between the performance of the ANOMA and that of the NOMA in terms of the throughput, although numerically it is shown that ANOMA outperforms NOMA in certain scenarios. This is probably because the existing expression for the throughput of the ANOMA system is given as a function of the channel matrix but not the signal-to-noise ratio (SNR). The lack of such an expression in terms of SNR makes the analytical comparison between NOMA and ANOMA almost intractable.  
Second, the optimal design of ANOMA has not been investigated. Despite the fact that the performance of ANOMA is directly affected by important design parameters such as the transmit power and the timing mismatch, existing papers mainly focused on the performance demonstration only.   
Third, the impact of timing error on the ANOMA systems has not been studied. Existing studies ideally assumed that the timing information was perfectly known. However, the timing information in practice cannot always be perfectly obtained, and the timing error is often inevitable~\cite{mostofi2006mathematical}. The timing information plays a vital role in ANOMA systems, since oversampling is designed using the timing information~\cite{xunGlobecom16,ganji2016interference}.

In this paper, we comprehensively investigate the ANOMA in a two-user uplink system. The primary contributions of the paper are summarized as follows:
\begin{enumerate}
	\item For the first time, we analytically prove that the ANOMA with a sufficiently large frame length can always outperform the NOMA in terms of the system sum throughput. To this end, we derive the expression for the sum throughput of the ANOMA system as a function of SNR, frame length, and normalized timing mismatch. A simplified throughput expression is further obtained for the asymptotic case of infinite frame length.
	
	
	\item We investigate the optimal design of the two-user uplink ANOMA system aiming at maximizing the sum throughput. We find that each user should transmit at full power despite the negative effect of inter-user interference.
	In addition, we prove that the optimal timing mismatch converges to one half of a symbol time as the frame length goes to infinity.
	
	\item We analyze the impact of timing error on the performance of the uplink ANOMA system. Two types of timing error are taken into consideration, i.e., the synchronization timing error and the coordination timing error, which account for the timing error caused in signal synchronization and the coordination of the timing mismatch between asynchronous signals, respectively. We derive the expressions for the throughput loss of the ANOMA system with respect to both types of timing error, and analyze how the synchronization timing error and the coordination timing error individually and jointly affect the system performance. 
\end{enumerate}

The remainder of the paper is organized as follows. The two-user uplink system model is presented in Section \ref{sectionSysModel}.
The performance of the ANOMA system is analyzed in Section \ref{sec:performanceAnalysisANOMA}. We discuss the optimal design of the ANOMA system in Section \ref{sectionSysdesign}.
We analyze the outputs of ANOMA matched filters with timing error and the throughput loss incurred by timing error in Section \ref{sectionImpactofError}. Numerical results are presented in Section \ref{sectionNumericalResults}. Finally, we draw the conclusions in Section \ref{sectionConclusion}.

Notations:
$(\cdot)^H$ denotes the Hermitian transpose, $(\cdot)^T$ denotes the transpose, $\Tr(\cdot)$ denotes the trace operation, $(\cdot)^{-1}$ denotes the inverse operation,
$|x|$ denotes the absolute value of $x$,
$\mathbb{E}[\cdot]$ denotes the expectation operation,
$\mathcal{CN}\left(0,1\right)$ denotes the complex normal distribution with zero mean and unit variance,
and $\mathbf{1}(\cdot)$ denotes the unit step function whose value is zero for negative arguments and one for positive arguments.

\section{System Model}\label{sectionSysModel}

\begin{figure}[t b]
	\centering
	\includegraphics[width=5in]{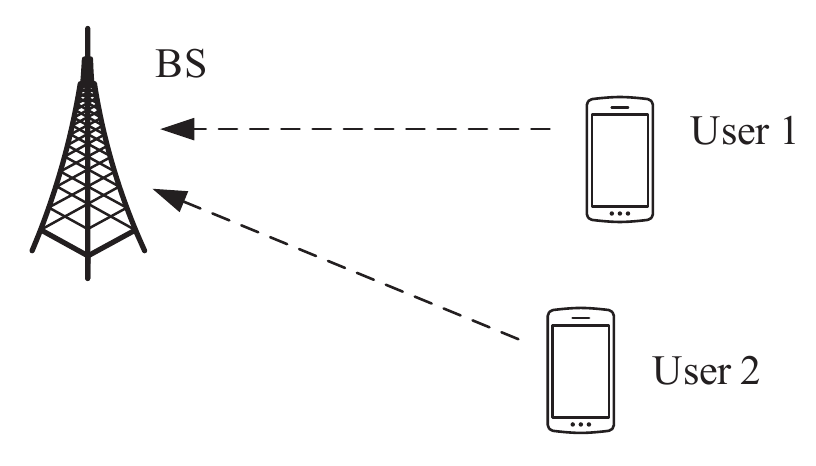}
	\caption{Illustration of a two-user uplink system.}
	\label{uplink}
\end{figure}

\begin{figure}[t b]
	\centering
	\includegraphics[width=5.5in]{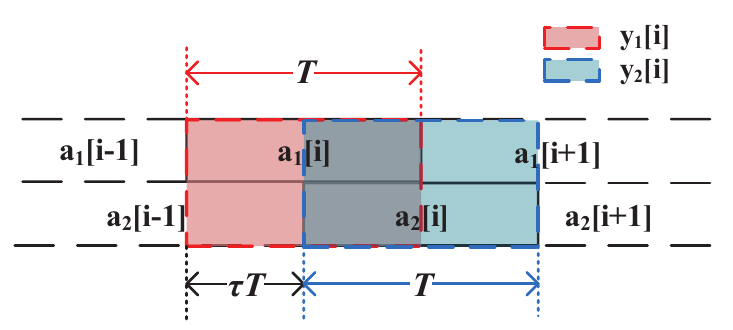}
	\caption{Illustration of the sampling for ANOMA.}
	\label{samplingnoerror}
\end{figure}

In this paper, we consider an uplink system which consists of a single base station (BS) and two users, as shown in Fig. \ref{uplink}. The two users share the same frequency-time resource to transmit signals to the BS. We assume that perfect channel state information (CSI) is known at the BS via the uplink channel training.

\subsection{ANOMA System}
For the ANOMA, a timing mismatch is intentionally introduced between the symbols from two users. As shown in Fig. \ref{samplingnoerror}, the intended timing mismatch between the symbols of Users 1 and 2 is denoted by $\tau T$, where $T$ is the duration of each symbol and $\tau$, $0 \le \tau < 1$, is the normalized timing mismatch. Note that the ANOMA system becomes a synchronous NOMA system when $\tau = 0$. In this section, we assume that $\tau$ is perfectly known at BS via timing offset estimation and uplink timing control techniques, such as the timing advance \cite{sesia2011lte}. We will study the ANOMA system with timing error in Section \ref{sectionImpactofError}.

Let $a_1[i] = h_1\sqrt{P_1}s_1[i]$ and $a_2[i]=h_2\sqrt{P_2}s_2[i]$, where
the subscripts $1$ and $2$ denote the parameters for Users 1 and 2, respectively,
$s_j[i]$ ($j=1,2$) denotes the $i$th normalized transmitted symbol, $h_j$ denotes the channel coefficient in the block of transmission, and $P_j$ denotes the transmit power. The BS's received signal at time $t$ is then given by
\begin{equation}\label{}
  y(t) = \sum_{i = 0}^{N-1} a_1[i]p(t - iT) + \sum_{i = 0}^{N-1} a_2[i]p(t - iT - \tau T) + n(t),
\end{equation}
where $N$ denotes the number of symbols in a frame, i.e., the frame length, $T$ denotes the time duration of one symbol, $p(\cdot)$ denotes the pulse-shaping filter with nonzero duration of $T$, and $n(t) \sim \mathcal{CN}(0, 1)$ denotes the normalized additive white Gaussian noise (AWGN). Without loss of generality, the rectangular pulse shape is adopted, i.e., $p(t)=1/T$ when $t \in [0, T]$ and $p(t)=0$ when $t \notin [0, T]$. With the block fading model, we assume that the channels remain static during the transmission of $N$ consecutive  symbols.

The BS uses the oversampling technique to take advantage of sampling diversity in the asynchronous systems~\cite{xunGlobecom16,xunICC18}. As shown in Fig. \ref{samplingnoerror}, the BS obtains two sample vectors, denoted by $[y_1[1], \cdots, y_1[N]]^T$ and $[y_2[1], \cdots, y_2[N]]^T$. Specifically, the $i$th element in the first sample vector is given by
\begin{align}\label{y_1i_noerror}
y_1[i] &= \int_{0}^{\infty}y(t)p(t - iT ) dt\notag\\
&= \int_{0}^{\infty}a_1[i] p(t-iT)p(t - iT ) dt\notag\\
& + \int_{0}^{\infty} \left\{a_2[i-1] p(t-(i+1+\tau) T) + a_2[i] p(t-(i+\tau) T)\right\}p(t - iT ) dt + n_1[i] \notag\\
&= a_1[i] + \tau a_2[i-1] + (1 - \tau) a_2[i] + n_1[i],
\end{align}
where $n_1[i]= \int_{0}^{\infty}n(t) p(t-iT)dt$ denotes the additive noise in the first sampled vector. The $i$th element in the second sample vector is given by
\begin{align}\label{y_2i_noerror}
y_2[i] &= \int_{0}^{\infty}y(t)p(t - iT -\tau T) dt = a_2[i] + \tau a_1[i+1] + (1 - \tau) a_1[i] + n_2[i],
\end{align}
where $n_2[i]=\int_{0}^{\infty}n(t) p(t-iT - \tau T)dt$ denotes the additive noise in the second sampled vector.

We can write the outputs at the BS in a matrix form as
\begin{align}\label{Y_noerror}
\mathbf{Y} = \mathbf{RHX} + \mathbf{N},
\end{align}
where
\begin{align}
\mathbf{Y} &= \left[y_1[1]\ y_2[1]\ y_1[2]\ y_2[2]\ \cdots\ y_1[N]\ y_2[N]\right]^T,\\
\mathbf{X} &= \left[s_1[1]\ s_2[1]\ s_1[2]\ s_2[2]\ \cdots\ s_1[N]\ s_2[N]\right]^T,\\
\mathbf{N} &= \left[n_1[1]\ n_2[1]\ n_1[2]\ n_2[2]\ \cdots\ n_1[N]\ n_2[N]\right]^T,\\
\mathbf{R} &= \left[\begin{smallmatrix}
	1\ &1-\tau\ &0\ &\cdots\ &\cdots\ &0\\
	1-\tau\ &1\ &\tau\ &0\ &\cdots\ &0\\
	0\ &\tau\ &1\ &1-\tau\ &\cdots\ &0\\
	\vdots\ &\ddots\ &\ddots\ &\ddots\ &\ddots\ &\vdots\\
	0\ &\cdots\ &0\ &\tau\ &1\ &1-\tau\\
	0\ &\cdots\ &\cdots\ &0\ &1-\tau\ &1
	\end{smallmatrix}\right],\label{R_matrix}
\end{align}
and
\begin{align}
\mathbf{H} &=
\left[\begin{smallmatrix}
h_1\sqrt{P_1} &       &       &       &\\
&h_2\sqrt{P_2} &       &       &\\
&       &\ddots &       &\\
&       &       &h_1\sqrt{P_1} &\\
&       &       &       &h_2\sqrt{P_2}
\end{smallmatrix}\right].\label{H_matrix}
\end{align}

We assume that the transmitted symbols are independent, such that $\mathbb{E}\left[\mathbf{XX}^H\right] = \mathbf{I}$. Note that the noise terms in \eqref{Y_noerror} are colored due to the oversampling, and we have
\begin{align}
\mathbb{E}\left\{n_1[i]n^H_2[i]\right\} &=
\int_{0}^{\infty}\int_{0}^{\infty}
\mathbb{E}\left\{n(t)n^H(s)\right\} p\left(t-iT\right)p\left(s-iT-\tau T\right) dt ds \notag\\
&= 1 - \tau.
\end{align}

Thus, the covariance matrix of $\mathbf{N}$ is given by
\begin{equation}\label{}
  \mathbf{R_{N}} = \mathbb{E}\left\{\mathbf{NN}^H\right\} = \mathbf{R}.
\end{equation}

\subsection{Benchmark System -- NOMA}
By setting $\tau = 0$, the ANOMA system becomes the synchronous NOMA system.  For the NOMA, the BS does not use the oversampling technique, and the $i$th sample at the BS can be written as
\begin{align}\label{y_noma}
y[i] = a_1[i] + a_2[i] + n[i],
\end{align}
where $n[i] = \int_{0}^{\infty} n(t)p(t-iT)dt$. Note that \eqref{y_noma} can be derived from either \eqref{y_1i_noerror} or \eqref{y_2i_noerror} by letting $\tau=0$.

\section{Performance Analysis of ANOMA Systems}\label{sec:performanceAnalysisANOMA}
In this section, we analyze the throughput performance of the ANOMA system. From \eqref{Y_noerror}, the sum throughput of the two-user uplink ANOMA system can be written as
\begin{align}\label{RsumANOMA}
R^{\mathrm{ANOMA}}= \frac{1}{N+\tau}\log \det\left(\mathbf{I}_{2N} + \mathbf{HH}^H\mathbf{R}\right).
\end{align}

We note that some existing papers define the throughput of ANOMA as
\begin{align}\label{RsumANOMA_ex}
R^{\mathrm{ANOMA}}_{\mathrm{exist}} = \frac{1}{N}\log \det\left(\mathbf{I}_{2N} + \mathbf{HH}^H\mathbf{R}\right),
\end{align}
which is slightly different from \eqref{RsumANOMA}. Although \eqref{RsumANOMA} and \eqref{RsumANOMA_ex} converge to the same expression as $N\rightarrow\infty$, we highlight that our adopted expression in \eqref{RsumANOMA} is more accurate than \eqref{RsumANOMA_ex} for evaluating the throughput of ANOMA with finite frame length $N$, since the system actually spends $N+\tau$ instead of $N$ symbol times to transmit $N$ symbols for each user.

It is worth mentioning that the practical transmission scheme may even simply allocate $N+1$ instead of $N+\tau$ symbol times for the transmission, and the throughput becomes $R_{N+1}^{\mathrm{ANOMA}}=\frac{1}{N+1}\log \det\left(\mathbf{I}_{2N} + \mathbf{HH}^H\mathbf{R}\right)$. Our analysis is still applicable to that case, since one can simply revise (most) results according to $R_{N+1}^{\mathrm{ANOMA}}= \frac{N + \tau}{N + 1} R^{\mathrm{ANOMA}}$.

In the following theorem, we derive the sum throughput of the two-user uplink ANOMA system in terms of the transmit SNRs, $\mu_1 = P_1|h_1|^2$ and $\mu_2 = P_2|h_2|^2$,  the normalized timing mismatch, $\tau$, and  the frame length, $N$.

\begin{Theorem}\label{theoremRsumANOMA}
	The sum throughput of the two-user uplink ANOMA system is derived as
	\begin{align}\label{eq:theoremRsumANOMA}
	R^{\mathrm{ANOMA}} = \frac{N}{N+\tau}\log\left(\mu_1\mu_2\right) + \frac{1}{N+\tau}\log \frac{(r_1^{N+1} - r_2^{N+1}) + \tau^2(r_1^N - r_2^N)}{r_1 - r_2},
	\end{align}
	where
	\begin{align}
	\mu_1 &= P_1|h_1|^2,\\
	\mu_2 &= P_2|h_2|^2,\\
	r_1 &=
	\frac{\mu_1^{-1}\! +\! \mu_2^{-1}\! +\! \mu_1^{-1}\mu_2^{-1}\! +\! 2\tau(1-\tau)\! +\! \sqrt{\left[\mu_1^{-1} + \mu_2^{-1} + \mu_1^{-1}\mu_2^{-1} + 2\tau(1-\tau)\right]^2\! -\! 4\tau^2(1-\tau)^2}}{2},\label{r_1}\\
	r_2 &=
		\frac{\mu_1^{-1}\! +\! \mu_2^{-1}\! +\! \mu_1^{-1}\mu_2^{-1}\! +\! 2\tau(1-\tau)\! -\! \sqrt{\left[\mu_1^{-1} + \mu_2^{-1} + \mu_1^{-1}\mu_2^{-1} + 2\tau(1-\tau)\right]^2\! -\! 4\tau^2(1-\tau)^2}}{2}.\label{r_2}
	\end{align}
\end{Theorem}
\begin{IEEEproof}
	See Appendix \ref{ProoftheoremRsumANOMA}.
\end{IEEEproof}

Based on Theorem~\ref{theoremRsumANOMA}, we present the throughput of the two-user uplink ANOMA system for the asymptotic case of $N\rightarrow\infty$ in the following corollary, which characterizes the limiting performance  of  the system when the frame length $N$ is large.

\begin{corollary}\label{theoremR_NtoInfinity}
	The throughput of the two-user uplink ANOMA system in the asymptotic case of $N\rightarrow\infty$ is given by
	\begin{equation}\label{}
	\lim_{N\rightarrow \infty} R^{\mathrm{ANOMA}} = \log\left(\mu_1\mu_2 r_1\right).
	\end{equation}	
\end{corollary}
\begin{IEEEproof}
	See Appendix \ref{ProoftheoremR_NtoInfinity}.
\end{IEEEproof}

\subsection{Comparison with NOMA}
According to \eqref{y_noma}, with perfect SIC at BS, the sum throughput of the two users in the uplink NOMA system can be written as \cite{goldsmith2003capacity}
\begin{align}\label{RsumNOMA}
R^{\mathrm{NOMA}} = \log (1 + P_1|h_1|^2 + P_2|h_2|^2) = \log (1 + \mu_1 + \mu_2),
\end{align}
which can also be obtained from \eqref{eq:theoremRsumANOMA} by setting $\tau=0$. 

Due to the complicated expression for the throughput of ANOMA in~\eqref{eq:theoremRsumANOMA}, it is difficult to analytically compare the NOMA with the ANOMA for a general value of $N$. Instead, we provide numerical results in Section \ref{sectionNumericalResults} and consider the asymptotic case of $N\rightarrow\infty$ for an analytical comparison in the following theorem.

\begin{Theorem}\label{theoremRanoma>Rnoma}
	The two-user uplink ANOMA system as $N\rightarrow\infty$ achieves an equal or higher throughput compared with the NOMA system, i.e.,
	\begin{align}
	\lim_{N\rightarrow \infty} R^{\mathrm{ANOMA}} \ge R^{\mathrm{NOMA}},
	\end{align}
	where $\lim_{N\rightarrow \infty} R^{\mathrm{ANOMA}} = R^{\mathrm{NOMA}}$ if and only if the normalized timing mismatch $\tau = 0$.
\end{Theorem}
\begin{IEEEproof}
	See Appendix \ref{ProofRanoma>Rnoma}.
\end{IEEEproof}

We further conclude the following corollary.
\begin{corollary}\label{corollarytau*0.5}
With a sufficiently large frame length, the ANOMA outperforms the NOMA for the two-user uplink system in terms of the sum throughput.
\end{corollary}

\section{Design of ANOMA Systems}\label{sectionSysdesign}
From the analysis in Section~\ref{sec:performanceAnalysisANOMA}, we note that the throughput performance of the uplink ANOMA system is directly determined by the transmit powers and the normalized timing mismatch, i.e., $P_1, P_2$, and $\tau$. In this section, we investigate the optimal $P_1, P_2$, and $\tau$ that maximize the throughput of the system.

The design problem is formulated as follows:
\begin{align}
 \mathop{\arg\max}_{P_1, P_2, \tau} &~~ R^{\mathrm{ANOMA}},\notag\\ 
 &s.t. ~~ 0\le\tau<1, ~ 0\le P_1\le  P_{1,\max}, ~ 0 \le P_2 \le P_{2,\max},
\end{align}
where $P_{1,\max}$ and $P_{2,\max}$ are the maximum available powers at which Users 1 and 2 can transmit, respectively.
Note that the transmit powers are coupled together in a complicated way in the expression for the throughput of the ANOMA system in \eqref{eq:theoremRsumANOMA}, which is different from the case of NOMA in~\eqref{RsumNOMA}.
Thus, the optimal transmit powers for the ANOMA system are not easy to determine, while it is easy to find that we shall use the maximum available transmit powers at users for the NOMA system.

It is worth mentioning that the performance of the uplink ANOMA system is also affected by the frame length, $N$. However, the frame length is constrained by the channel condition, i.e., the length of each block of the block fading channel, and the acceptable transceiver complexity. Hence, we do not investigate the design of $N$ in this work and assume that it is fixed based on the channel conditions and the overall system design.

\subsection{Optimal Transmit Power}
We first obtain the optimal transmit power scheme. We summarize the optimal transmit powers for the ANOMA system as follows.

\begin{Theorem}\label{FullPowerTheorem}
For the two-user uplink ANOMA system with any frame length, $N$, and the normalized timing mismatch, $\tau$, the optimal transmit powers at Users 1 and 2, $P_1^*$ and $P_2^*$, are equal to the maximum available powers at which Users 1 and 2 can transmit, $P_{1,\max}$ and $P_{2,\max}$, i.e., $P_1^*=P_{1,\max}$ and $P_2^*=P_{2,\max}$
\end{Theorem}
\begin{IEEEproof}
	See Appendix \ref{ProofFullPowerTheorem}.
\end{IEEEproof}

From Theorem \ref{FullPowerTheorem}, we find that the optimal design of transmit powers for the two-user uplink ANOMA system is the same as that for the NOMA system.

\subsection{Optimal Normalized Timing Mismatch}
We now study the optimal normalized timing mismatch, $\tau^*$. The optimal normalized timing mismatch, $\tau^*$, is analytically intractable for a general finite frame length $N$, while we can numerically obtain $\tau^*$ for a given finite $N$ by simply searching in the range of $0\le\tau<1$.
In addition, we present $\tau^*$ in the asymptotic case of $N\rightarrow\infty$ in the following theorem.

\begin{Theorem}\label{theoremtau*0.5}
	For the two-user uplink ANOMA system with the frame length $N\rightarrow\infty$, the optimal normalized timing mismatch to maximize the sum throughput is given by $\tau^*=0.5$.
\end{Theorem}
\begin{IEEEproof}
	See Appendix \ref{prooftheoremtau*0.5}.
\end{IEEEproof}


\section{Impact of Timing Error on ANOMA systems}\label{sectionImpactofError}
The analysis in the previous sections is based on the assumption that the BS perfectly knows the timing information. However, the timing information cannot always be perfectly obtained in practice, and the timing error is often inevitable. In this section, we study the impact of timing error on the ANOMA system.

\subsection{Timing Error}
We consider two types of timing error for the ANOMA system, i.e., the synchronization timing error and the coordination timing error.

\subsubsection{Synchronization Timing Error}\label{sec:tsr}
To synchronize the signals, we need a reference signal. Without loss of generality, we use the signal from User 1 as the timing reference (the timing offset is 0). This requires a symbol-level timing synchronization with User 1 at the BS, as it is also done in NOMA.
The normalized synchronization timing error, denoted by $\epsilon_1$ in Fig. \ref{samplingerror}, is due to the imperfect timing synchronization.
Without loss of generality, we assume that $\epsilon_1 \in (\tau - 1, \tau)$.
With the synchronization timing error,  $y_1[i]$ is taken from the time $(i-1)T+\epsilon_1 T$ to $iT+\epsilon_1 T$ and $y_2[i]$ is taken from the time $(i-1)T+(\tau +\epsilon_1)T$ to $iT+(\tau +\epsilon_1)T$, although the BS intends to take $y_1[i]$ from the time $(i - 1)T$ to $iT$ and $y_2[i]$  from the time $(i-1)T+\tau T$ to $iT+\tau T$. We will study the effect of this timing error later.

\subsubsection{Coordination Timing Error}
\begin{figure}[t]
	\centering
	\includegraphics[width=5.5in]{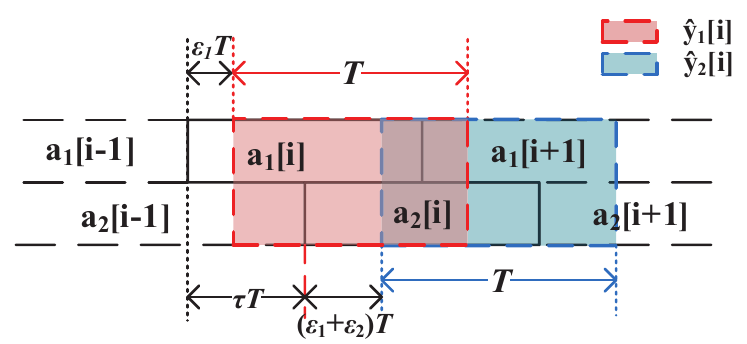}
	\caption{Illustration of the sampling for ANOMA with timing error.}
	\label{samplingerror}
\end{figure}

In order to achieve the desired timing mismatch between the two signals, the BS coordinates the uplink transmission timing of the two users to add the intended timing offsets at each transmitter. For example, the timing advance is the technique employed in long term evolution (LTE) systems to estimate and adjust the timing offsets among uplink signals at BS \cite{access2011physical,sesia2011lte}. The normalized coordination timing error, denoted by $\epsilon_2$ in Fig. \ref{samplingerror}, results from the imperfect coordination between the users. With the coordination timing error, the actual timing mismatch becomes $(\tau + \epsilon_2) T$, while the intended timing mismatch is $\tau T$. In addition to the synchronization timing error $\epsilon_1 T$, the sample $y_2[i]$ is taken from $(i-1)T + (\tau + \epsilon_1 + \epsilon_2)T$ to $iT + (\tau + \epsilon_1 + \epsilon_2)T$, although the BS intends to take $y_2[i]$ from $(i - 1)T + \tau T$ to $iT + \tau T$. Without loss of generality, we assume that $\epsilon_1 + \epsilon_2 \in (-\tau , 1 - \tau)$.

Fig.~\ref{samplingerror} illustrates the sampling for an ANOMA system with timing error. It is worth mentioning that the sign of the timing error stands for the direction in which the function of the matched filter is shifted. For example, as shown in Fig. \ref{samplingerror}, the matched filter is shifted to the right by $\epsilon_1 T$ if $\epsilon_1 > 0$ compared to the matched filter designed with no timing error in Fig. \ref{samplingnoerror}. Fig. \ref{samplingerror} only presents the case when $\epsilon_1 > 0$ and $\epsilon_1 + \epsilon_2 > 0$, while our analysis works for any values of $\epsilon_1$ and $\epsilon_2$.

\subsection{Outputs of ANOMA Matched Filters with Timing Error}
We now derive the outputs of the matched filters at the BS with both the synchronization timing error and the coordination timing error.

In the presence of timing error, the $i$th element of the first sample vector is given by
\begin{align}\label{y1i}
\hat{y}_1[i] &= \int_{0}^{\infty} y(t) p(t - iT - \epsilon_1 T)dt \notag\\
&= \int_{0}^{\infty} a_1[i]p(t - iT)p(t - iT - \epsilon_1 T)dt \notag\\
&+ \mathbf{1}(-\epsilon_1) \int_{0}^{\infty} a_1[i-1]p(t - (i-1)T)p(t - iT - \epsilon_1 T)dt \notag\\
&+ \mathbf{1}(\epsilon_1) \int_{0}^{\infty} a_1[i+1]p(t - (i+1)T)p(t - iT - \epsilon_1 T)dt \notag\\
&+ \int_{0}^{\infty} a_2[i-1]p(t - \tau T - (i-1)T) p(t-iT-\epsilon_1 T)dt \notag\\
&+ \int_{0}^{\infty} a_2[i]p(t - \tau T - iT) p(t-iT-\epsilon_1T)dt + \int_{0}^{\infty} n(t)p(t-iT-\epsilon_1T)dt \notag\\
&= a_1[i] (1 - |\epsilon_1|) + a_1[i-1]\mathbf{1}(-\epsilon_1) (-\epsilon_1) + a_1[i+1]\mathbf{1}(\epsilon_1)\epsilon_1 \notag\\
&+ a_2[i-1](\tau - \epsilon_1) + a_2[i](1-\tau+\epsilon_1) + \hat{n}_1[i],
\end{align}
and
the $i$th element of the second sample vector is given by
\begin{align}\label{y2i}
\hat{y}_2[i] &= \int_{0}^{\infty}y(t)p(t - (i + \tau + \epsilon_1 + \epsilon_2)T) dt \notag\\
& = a_2[i] (1 - |\epsilon_1+\epsilon_2|) + a_2[i-1]\mathbf{1}(-\epsilon_1- \epsilon_2) (-\epsilon_1-\epsilon_2) \notag\\
&+ a_2[i+1]\mathbf{1}(\epsilon_1 + \epsilon_2)(\epsilon_1 + \epsilon_2) + a_1[i](\tau - \epsilon_1 - \epsilon_2) + a_1[i+1](1-\tau + \epsilon_1 + \epsilon_2) + \hat{n}_2[i],
\end{align}
where $\hat{n}_1[i] = \int_{0}^{\infty} n(t)p(t-iT-\epsilon_1T)dt$ and $\hat{n}_2[i] = \int_{0}^{\infty} n(t)p(t- (i + \tau + \epsilon_1 + \epsilon_2)T)dt$.

We note from \eqref{y1i} and \eqref{y2i} that the first sample vector is affected by the normalized synchronization timing error $\epsilon_1$ only, while the second sample vector is affected by the sum of the normalized synchronization timing error $\epsilon_1$ and the normalized coordination timing error $\epsilon_2$. 

With \eqref{y1i} and \eqref{y2i}, we obtain the outputs of the two matched filters at the BS subject to the timing error in
the matrix form as
\begin{equation}
\mathbf{\hat{Y}} = \mathbf{\hat{R}HX} + \mathbf{\hat{N}},
\label{Y_error}
\end{equation}
where
$\hat{\mathbf{Y}} = [\hat{y}_1[1]\ \hat{y}_2[1]\ \hat{y}_1[2]\ \hat{y}_2[2]\ \cdots\ \hat{y}_1[N]\ \hat{y}_2[N]]^T$,
$\mathbf{\hat{N}} = [\hat{n}_1[1]\ \hat{n}_2[1]\ \hat{n}_1[2]\ \hat{n}_2[2]\ \cdots\ \hat{n}_1[N]$ $\hat{n}_2[N]]^T$, and  $\hat{\mathbf{R}}$ is given by
\begin{align}\label{wideeq}
	&\mathbf{\hat{R}}\notag\\
	&= \left[\begin{smallmatrix}
	1-|\epsilon_1|\ &1-\tau+\epsilon_1\ &\mathbf{1}(\epsilon_1)\epsilon_1\ &0\ &\cdots\ &\cdots\ &0\\
	1-\tau-\epsilon_1-\epsilon_2\ &1-|\epsilon_1+\epsilon_2|\ &\tau+\epsilon_1+\epsilon_2\ &\mathbf{1}(\epsilon_1+\epsilon_2)(\epsilon_1+\epsilon_2)\ &0\ &\cdots\ &0\\
	\mathbf{1}(-\epsilon_1)(-\epsilon_1)\ &\tau-\epsilon_1\ &1-|\epsilon_1|\ &1-\tau+\epsilon_1\ &\mathbf{1}(\epsilon_1)\epsilon_1\ &\cdots\ &0\\
	\vdots\ &\ddots\ &\ddots\ &\ddots\ &\ddots\ &\ddots\ &\vdots\\
	0\ &\cdots\ &\mathbf{1}(-\epsilon_1-\epsilon_2)(-\epsilon_1-\epsilon_2) &1-\tau-\epsilon_1-\epsilon_2\ &1-|\epsilon_1+\epsilon_2|\ &\tau+\epsilon_1+\epsilon_2\ &\mathbf{1}(\epsilon_1+\epsilon_2)(\epsilon_1+\epsilon_2)\\
	0\ &\cdots\ &0\ &\mathbf{1}(-\epsilon_1)(-\epsilon_1)\ &\tau-\epsilon_1\ &1-|\epsilon_1|\ &1-\tau+\epsilon_1\\
	0\ &\cdots\ &\cdots\ &0\ &\mathbf{1}(-\epsilon_1-\epsilon_2)(-\epsilon_1-\epsilon_2) &1-\tau-\epsilon_1-\epsilon_2\ &1-|\epsilon_1+\epsilon_2|\end{smallmatrix}\right] \notag\\
	&= \mathbf{R}\notag\\ 
	&+ \underbrace{\left[\begin{smallmatrix}
		-|\epsilon_1|\ &\epsilon_1\ &\mathbf{1}(\epsilon_1)\epsilon_1\ &0\ &\cdots\ &\cdots\ &0\\
		-\epsilon_1-\epsilon_2\ &-|\epsilon_1+\epsilon_2|\ &\epsilon_1+\epsilon_2\ &\mathbf{1}(\epsilon_1+\epsilon_2)(\epsilon_1+\epsilon_2)\ &0\ &\cdots\ &0\\
		\mathbf{1}(-\epsilon_1)(-\epsilon_1)\ &-\epsilon_1\ &-|\epsilon_1|\ &\epsilon_1\ &\mathbf{1}(\epsilon_1)\epsilon_1\ &\cdots\ &0\\
		\vdots\ &\ddots\ &\ddots\ &\ddots\ &\ddots\ &\ddots\ &\vdots\\
		0\ &\cdots\ &\mathbf{1}(-\epsilon_1-\epsilon_2)(-\epsilon_1-\epsilon_2) &-\epsilon_1-\epsilon_2\ &-|\epsilon_1+\epsilon_2|\ &\epsilon_1+\epsilon_2\ &\mathbf{1}(\epsilon_1+\epsilon_2)(\epsilon_1+\epsilon_2)\\
		0\ &\cdots\ &0\ &\mathbf{1}(-\epsilon_1)(-\epsilon_1)\ &-\epsilon_1\ &-|\epsilon_1|\ &\epsilon_1\\
		0\ &\cdots\ &\cdots\ &0\ &\mathbf{1}(-\epsilon_1-\epsilon_2)(-\epsilon_1-\epsilon_2) &-\epsilon_1-\epsilon_2\ &-|\epsilon_1+\epsilon_2|\end{smallmatrix}\right]}_{\mathbf{E}_1}.
\end{align}

We note from \eqref{wideeq} that the expression for $\mathbf{E}_1$ is related to the signs of $\epsilon_1$ and $\epsilon_1 + \epsilon_2$. 
For the sake of brevity, we present the analytical results for the case of $\epsilon_1 > 0$ and $\epsilon_1 + \epsilon_2 > 0$ in the rest of the paper, while our analytical method and findings are applicable to all cases.
In addition, we will present the numerical results in Section \ref{sectionNumericalResults} for all possible cases of $\epsilon_1$ and $\epsilon_1 + \epsilon_2$.
With $\epsilon_1 > 0$ and $\epsilon_1 + \epsilon_2 >0$, the expression for $\mathbf{E}_1$ is rewritten as
\begin{align}\label{E_1}
\mathbf{E}_1 &= \epsilon_1\underbrace{\left[\begin{smallmatrix}
	-1\ &1\ &1\ &0\ &\cdots\ &\cdots\ &0\\
	-1\ &-1\ &1\ &1\ &0\ &\cdots\ &0\\
	0\ &-1\ &-1\ &1\ &1\ &\cdots\ &0\\
	\vdots\ &\ddots\ &\ddots\ &\ddots\ &\ddots\ &\ddots\ &\vdots\\
	0\ &\cdots\ &0 &-1\ &-1\ &1\ &1\\
	0\ &\cdots\ &0\ &0\ &-1\ &-1\ &1\\
	0\ &\cdots\ &\cdots\ &0\ &0 &-1\ &-1\end{smallmatrix}\right]}_{\mathbf{Z}_1} + \epsilon_2\underbrace{\left[\begin{smallmatrix}
	0\ &0\ &0\ &0\ &\cdots\ &\cdots\ &0\\
	-1\ &-1\ &1\ &1\ &0\ &\cdots\ &0\\
	0\ &0\ &0\ &0\ &0\ &\cdots\ &0\\
	\vdots\ &\ddots\ &\ddots\ &\ddots\ &\ddots\ &\ddots\ &\vdots\\
	0\ &\cdots\ &0 &-1\ &-1\ &1\ &1\\
	0\ &\cdots\ &0\ &0\ &0\ &0\ &0\\
	0\ &\cdots\ &\cdots\ &0\ &0 &-1\ &-1\end{smallmatrix}\right]}_{\mathbf{Z}_2}.
\end{align}

The covariance matrix of $ \mathbf{\hat{N}}$ is given by
\begin{align}\label{RNN}
\mathbf{\hat{R}_{N}} &= \mathbb{E}\left\{\mathbf{\hat{N}\hat{N}}^H\right\} \notag\\
&= \left[\begin{smallmatrix}
1\ &1-\tau-\epsilon_2\ &0\ &\cdots\ &\cdots\ &0\\
1-\tau-\epsilon_2\ &1\ &\tau+\epsilon_2\ &0\ &\cdots\ &0\\
0\ &\tau+\epsilon_2\ &1\ &1-\tau-\epsilon_2\ &\cdots\ &0\\
\vdots\ &\ddots\ &\ddots\ &\ddots\ &\ddots\ &\vdots\\
0\ &\cdots\ &0\ &\tau+\epsilon_2\ &1\ &1-\tau-\epsilon_2\\
0\ &\cdots\ &\cdots\ &0\ &1-\tau-\epsilon_2\ &1
\end{smallmatrix}\right]\notag\\
&= \mathbf{R} + \underbrace{\left[\begin{smallmatrix}
	0\ &-\epsilon_2 \ &0\ &\cdots\ &\cdots\ &0\\
	-\epsilon_2\ &0\ &\epsilon_2\ &0\ &\cdots\ &0\\
	0\ &\epsilon_2\ &0\ &-\epsilon_2\ &\cdots\ &0\\
	\vdots\ &\ddots\ &\ddots\ &\ddots\ &\ddots\ &\vdots\\
	0\ &\cdots\ &0\ &\epsilon_2\ &0\ &-\epsilon_2\\
	0\ &\cdots\ &\cdots\ &0\ &-\epsilon_2\ &0
	\end{smallmatrix}\right]}_{\mathbf{E}_2},
\end{align}
where $\mathbf{E}_2$ can be rewritten as
\begin{equation}\label{E2}
\mathbf{E}_2 = \epsilon_2 \underbrace{\left[\begin{smallmatrix}
0\ &-1 \ &0\ &\cdots\ &\cdots\ &0\\
-1\ &0\ &1\ &0\ &\cdots\ &0\\
0\ &1\ &0\ &-1\ &\cdots\ &0\\
\vdots\ &\ddots\ &\ddots\ &\ddots\ &\ddots\ &\vdots\\
0\ &\cdots\ &0\ &1\ &0\ &-1\\
0\ &\cdots\ &\cdots\ &0\ &-1\ &0
\end{smallmatrix}\right]}_{\mathbf{Z}_3}.
\end{equation}

We note from \eqref{E2} that the covariance matrix of the noise terms
is affected by the normalized coordination timing error $\epsilon_2$, while it is not related to the normalized synchronization timing error $\epsilon_1$.

\subsection{Impact of Timing Error on Throughput Performance}
According to (\ref{Y_error}), the throughput of the ANOMA system with timing error is given by
\begin{align}\label{R_e}
	&R_e^{\mathrm{ANOMA}} \notag\\
	&= \frac{1}{N+\tau}\log\det\left(\mathbf{I}_{2N} + \mathbf{\hat{R}_{N}}^{-1}\mathbf{\hat{R}HH}^H\mathbf{\hat{R}}^H\right)\notag\\
	&= \frac{1}{N+\tau}\log\det\left(\mathbf{I}_{2N} + (\mathbf{R+E}_2)^{-1}(\mathbf{R + E}_1)\mathbf{HH}^H(\mathbf{R + E}_1^H)\right)\notag\\
	&= \frac{1}{N+\tau}\log\det\left(\mathbf{I}_{2N} + \left(\mathbf{I}_{2N} + (\mathbf{R+E}_2)^{-1}(\mathbf{E}_1 - \mathbf{E}_2)\right) \mathbf{HH}^H(\mathbf{R + E}_1^H)\right)\notag\\
	&= \frac{1}{N+\tau}\log\det\left(\mathbf{I}_{2N}\! +\! \mathbf{HH}^H\mathbf{R} + \mathbf{HH}^H\mathbf{E}_1^H \! +\! (\mathbf{R + E}_2)^{-1}(\mathbf{E}_1 - \mathbf{E}_2)\mathbf{HH}^H(\mathbf{R\! +\! E}_1^H)\right).
\end{align}

When there is no timing error, i.e., $\epsilon_1=\epsilon_2=0$, we have $\mathbf{E}_1 = \mathbf{E}_2=\mathbf{0}$. Hence, substituting $\mathbf{E}_1 = \mathbf{E}_2=\mathbf{0}$ into \eqref{R_e}, we obtain the throughput of the ANOMA system without timing error, which is the same as \eqref{RsumANOMA}.

From \eqref{RsumANOMA} and \eqref{R_e}, we derive the throughput loss incurred by the timing error as
\begin{align}\label{rate_diff}
	\Delta &= R^{\mathrm{ANOMA}} - R_e^{\mathrm{ANOMA}} \notag\\
	&= -\frac{1}{N + \tau}\log\det\left\{\mathbf{I}_{2N} + \left(\mathbf{I}_{2N} + \mathbf{HH}^H\mathbf{R}\right)^{-1}\left[\mathbf{HH}^H\mathbf{E}_1^H \right.\right.\notag\\
	&\left.\left. + (\mathbf{R + E}_2)^{-1}(\mathbf{E}_1 - \mathbf{E}_2)\mathbf{HH}^H(\mathbf{R + E}_1^H)\right] \right\}.
\end{align}

In what follows, we separately analyze the throughput loss incurred by the synchronization timing error and the coordination timing error with the practical consideration that these two types of timing error both are relatively small. For each case, we show that the throughput loss is approximately linear to the timing error by omitting the high-order terms of the timing error.

\subsubsection{Impact of Synchronization Timing Error}
We first investigate the impact of synchronization timing error on the throughput loss and consider the practical scenario where the error is relatively small, such that $\epsilon_2 = 0$ and $\epsilon_1\ll1$.

In this case, by omitting high-order terms of $\epsilon_1$, we obtain the throughput loss incurred by the synchronization timing error from \eqref{rate_diff} as
\begin{align}\label{delta_e1}
\Delta_{\epsilon_1} &= -\frac{1}{N\!+\!\tau}\log\det\left\{\mathbf{I}_{2N}\! +\! \epsilon_1\left(\mathbf{I}_{2N}\! +\! \mathbf{HH}^H\mathbf{R}\right)^{-1}\left[\mathbf{HH}^H\mathbf{Z}_1^H \!+\! \mathbf{R}^{-1}\mathbf{Z}_1\mathbf{HH}^H(\mathbf{R}\! +\! \epsilon_1\mathbf{Z}_1^H)\right] \right\} \notag\\
&\stackrel{(a)}{\approx} -\frac{1}{N+\tau}\log\det\left\{\mathbf{I}_{2N} + \epsilon_1\left(\mathbf{I}_{2N} + \mathbf{HH}^H\mathbf{R}\right)^{-1}\left[\mathbf{HH}^H\mathbf{Z}_1^H + \mathbf{R}^{-1}\mathbf{Z}_1\mathbf{HH}^H\mathbf{R}\right] \right\} \notag\\
&\stackrel{(b)}{\approx} -\frac{1}{N+\tau}\log\left( 1 + \epsilon_1\Tr(\mathbf{F}_1) + O(\epsilon_1^2)\right) \notag\\
&\stackrel{(c)}{\approx} \epsilon_1 c_1,
\end{align}
where $\mathbf{F}_1 = \left(\mathbf{I}_{2N} + \mathbf{HH}^H\mathbf{R}\right)^{-1}\left(\mathbf{HH}^H\mathbf{Z}_1^H +  \mathbf{R}^{-1}\mathbf{Z}_1\mathbf{HH}^H\mathbf{R}\right)$, $c_1 = -\frac{1}{N+\tau}\Tr(\mathbf{F}_1)$, $(a)$ is approximated by using $\mathbf{R} + \epsilon_1 \mathbf{Z} \approx \mathbf{R}$ as $\epsilon_1 \rightarrow 0$, $(b)$ is derived using the special case of Jacobi's formula \cite{magnus1988matrix}, i.e., $\det\left(\mathbf{I} + \epsilon\mathbf{A}\right) = 1 + \epsilon\Tr(\mathbf{A}) + O(\epsilon^2)$, and $(c)$ is derived by omitting the high-order terms of $\epsilon_1$ and applying the approximation $\log(1 + x) \approx x$ when $x \ll 1$. From  \eqref{delta_e1}, we note that the throughput loss is approximately linear to $\epsilon_1$ when $\epsilon_2 = 0$ and $\epsilon_1\ll1$.


\subsubsection{Impact of Coordination Timing Error}
We now investigate the impact of the coordination timing error on the throughput loss and still consider the practical scenario where the error is relatively small, such that $\epsilon_1 = 0$ and $\epsilon_2 \ll 1$.

By omitting high-order terms of $\epsilon_2$, we obtain the throughput loss incurred by the coordination timing error from \eqref{rate_diff}~as
\begin{align}\label{delta_e2}
\Delta_{\epsilon_2} &= -\frac{1}{N+\tau}\log\det\left\{\mathbf{I}_{2N} + \epsilon_2\left(\mathbf{I}_{2N} + \mathbf{HH}^H\mathbf{R}\right)^{-1} \left[\mathbf{HH}^H\mathbf{Z}_2^H \right.\right.\notag\\
&\left.\left. +(\mathbf{R} + \epsilon_2\mathbf{Z}_3)^{-1}(\mathbf{Z}_2 - \mathbf{Z}_3)\mathbf{HH}^H(\mathbf{R} +\epsilon_2\mathbf{Z}_2^H)\right] \right\}\notag\\
&\stackrel{(a)}{\approx} \epsilon_2 c_2,
\end{align}
where $\mathbf{F}_2 = \left(\mathbf{I}_{2N} + \mathbf{HH}^H\mathbf{R}\right)^{-1}\left(\mathbf{HH}^H\mathbf{Z}_2^H + \mathbf{R}^{-1} (\mathbf{Z}_2 - \mathbf{Z}_3)\mathbf{HH}^H\mathbf{R}\right)$, $c_2 = -\frac{1}{N+\tau}\Tr(\mathbf{F}_2)$, and $(a)$ can be derived by following the same steps in the derivation of \eqref{delta_e1}.
From \eqref{delta_e2}, we note that the throughput loss is approximately linear to $\epsilon_2$ when $\epsilon_1 = 0$ and $\epsilon_2\ll1$.

\section{Numerical Results}\label{sectionNumericalResults}

\begin{figure}[t b]
	\centering
	\includegraphics[width=0.9\columnwidth]{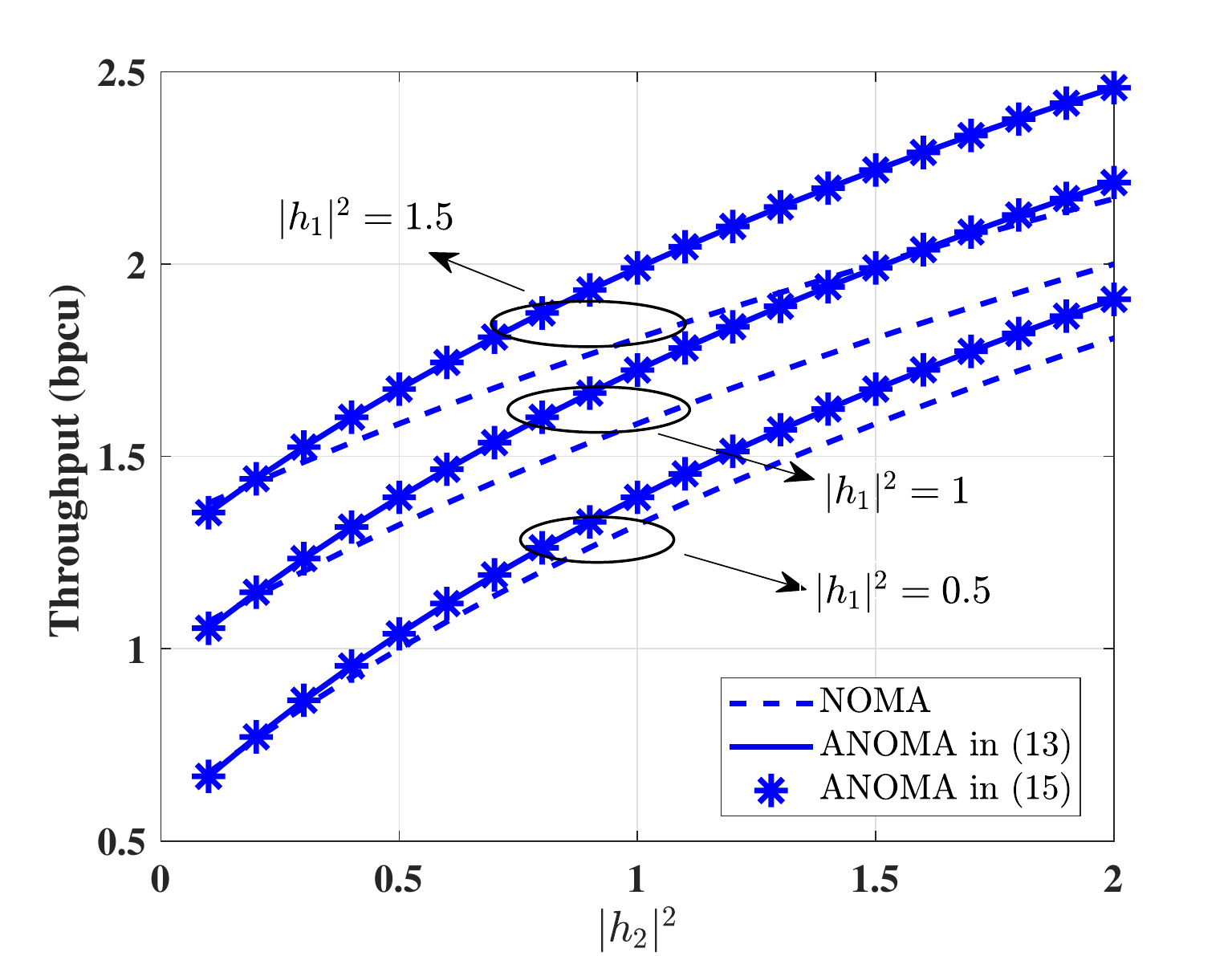}
	\caption{The sum throughput of two users as a function of channel gains for ANOMA and NOMA systems when $P_1 = 1$, $P_2 = 1$, $\tau = 0.5$, and $N = 10$.}
	\label{rate_vs_h}
\end{figure}
\begin{figure}[t b]
	\centering
	\includegraphics[width=0.9\columnwidth]{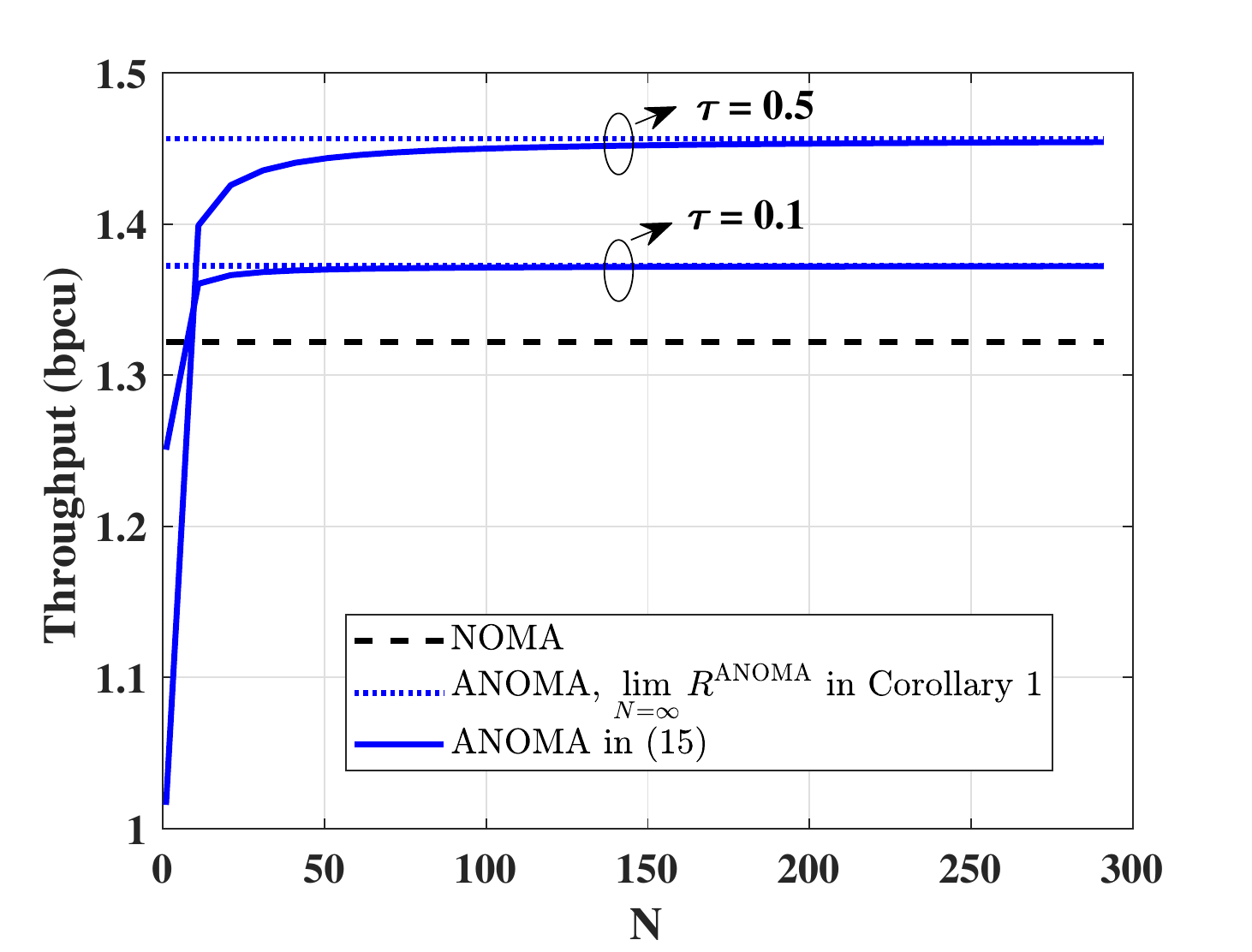}
	\caption{The sum throughput of two users as a function of the frame length $N$ for ANOMA and NOMA systems when $P_1|h_1|^2 = 1$, $P_2|h_2|^2 = 0.5$, $\tau = 0.5$ or 0.1.}
	\label{compareANOMAandNOMA}
\end{figure}
\begin{figure}[t b]
	\centering
	\includegraphics[width=0.9\columnwidth]{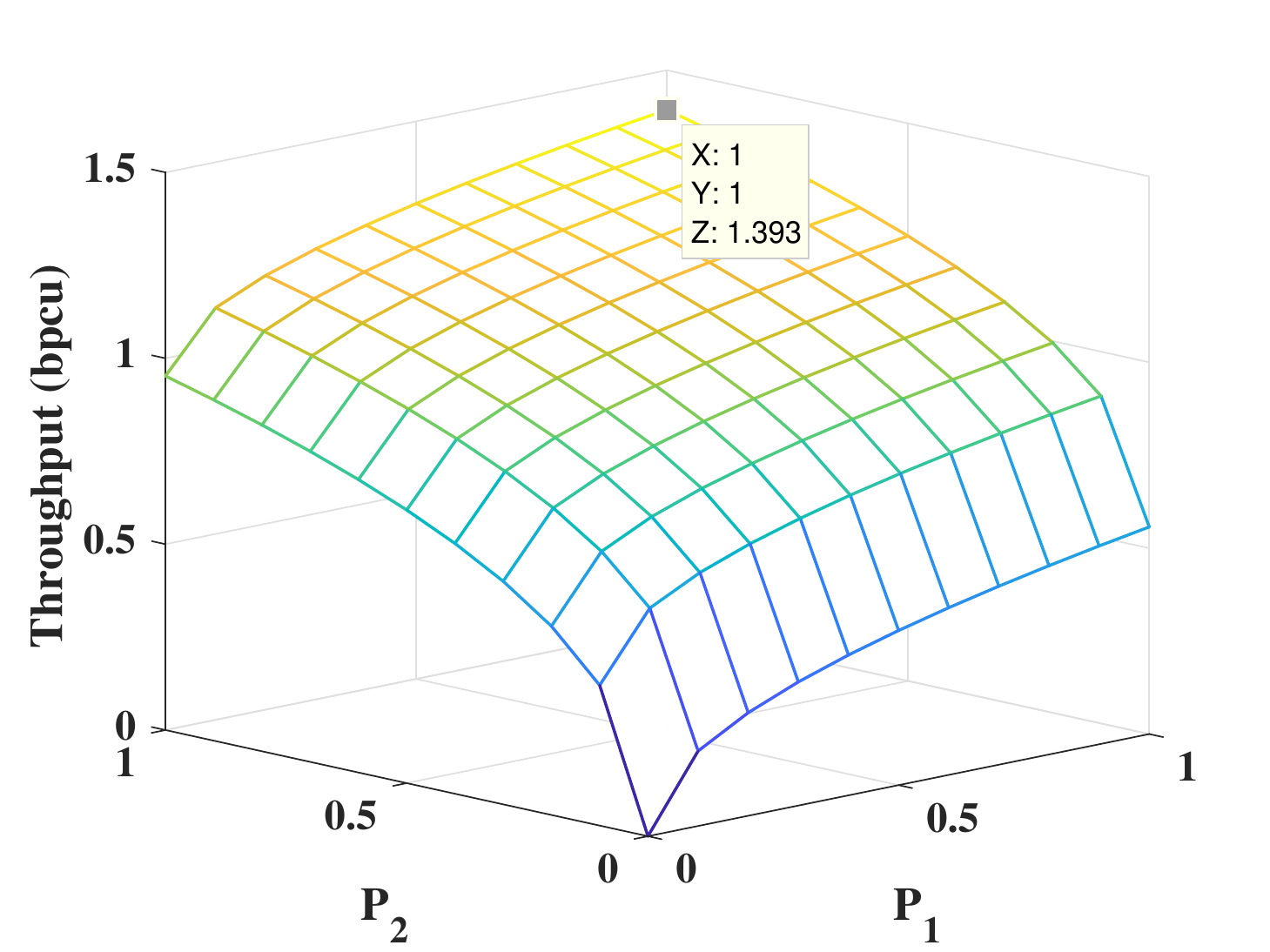}
	\caption{The sum throughput of two users as a function of transmit powers of Users 1 \& 2 for ANOMA systems when $|h_1|^2 = 1$, $|h_2|^2 = 0.5$, $P_{1,\mathrm{max}}=P_{2,\mathrm{max}}=1$, $\tau = 0.5$, and $N = 10$.}
	\label{maxPowerTransmit}
\end{figure}
\begin{figure}[t b]
	\centering
	\includegraphics[width=0.9\columnwidth]{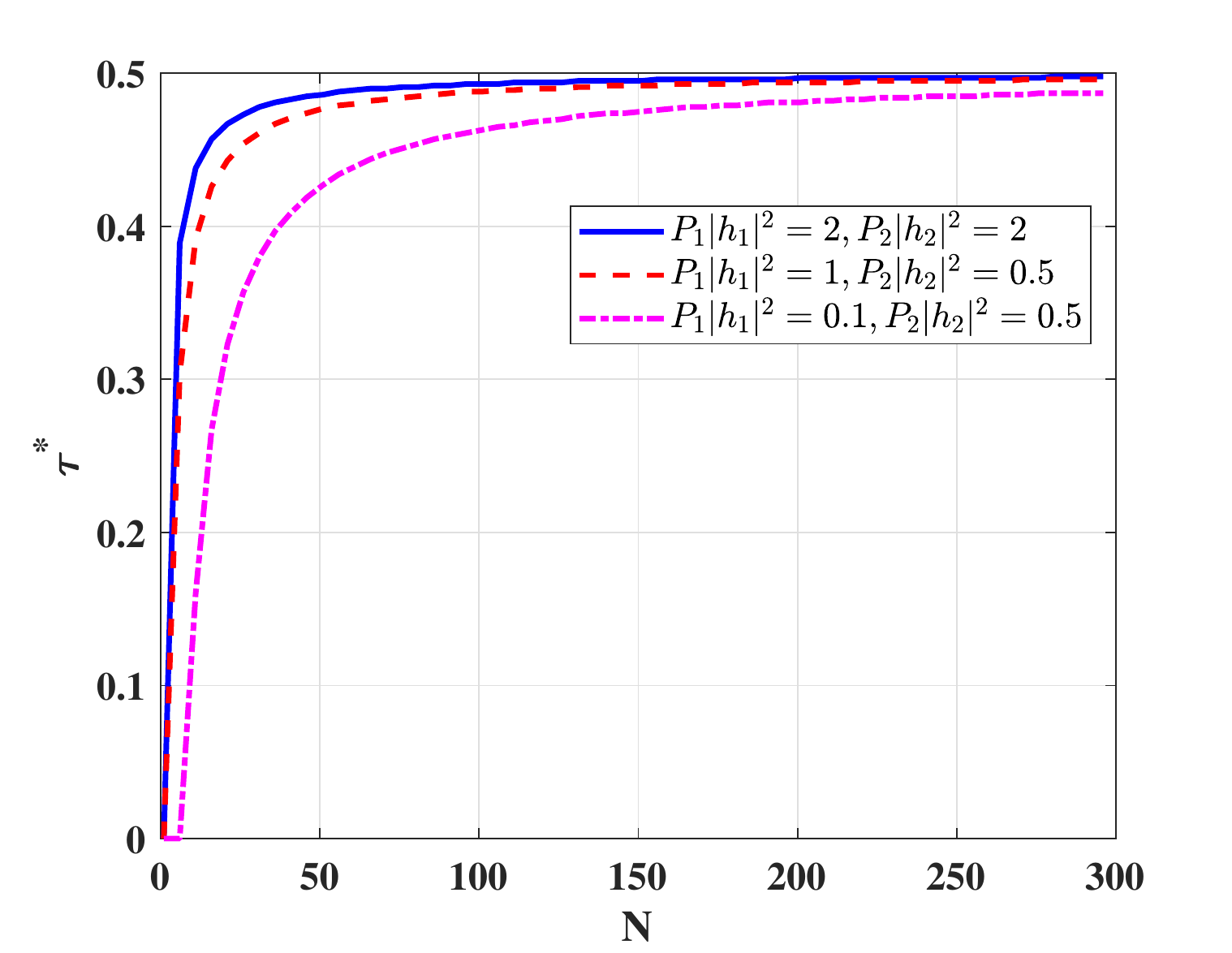}
	\caption{The optimal normalized timing mismatch $\tau^*$ to maximize the sum throughput of two users as a function of the frame length $N$ for different channel conditions.}
	\label{optimal_tau}
\end{figure}

\begin{figure}[t b]
	\centering
	\includegraphics[width=0.9\columnwidth]{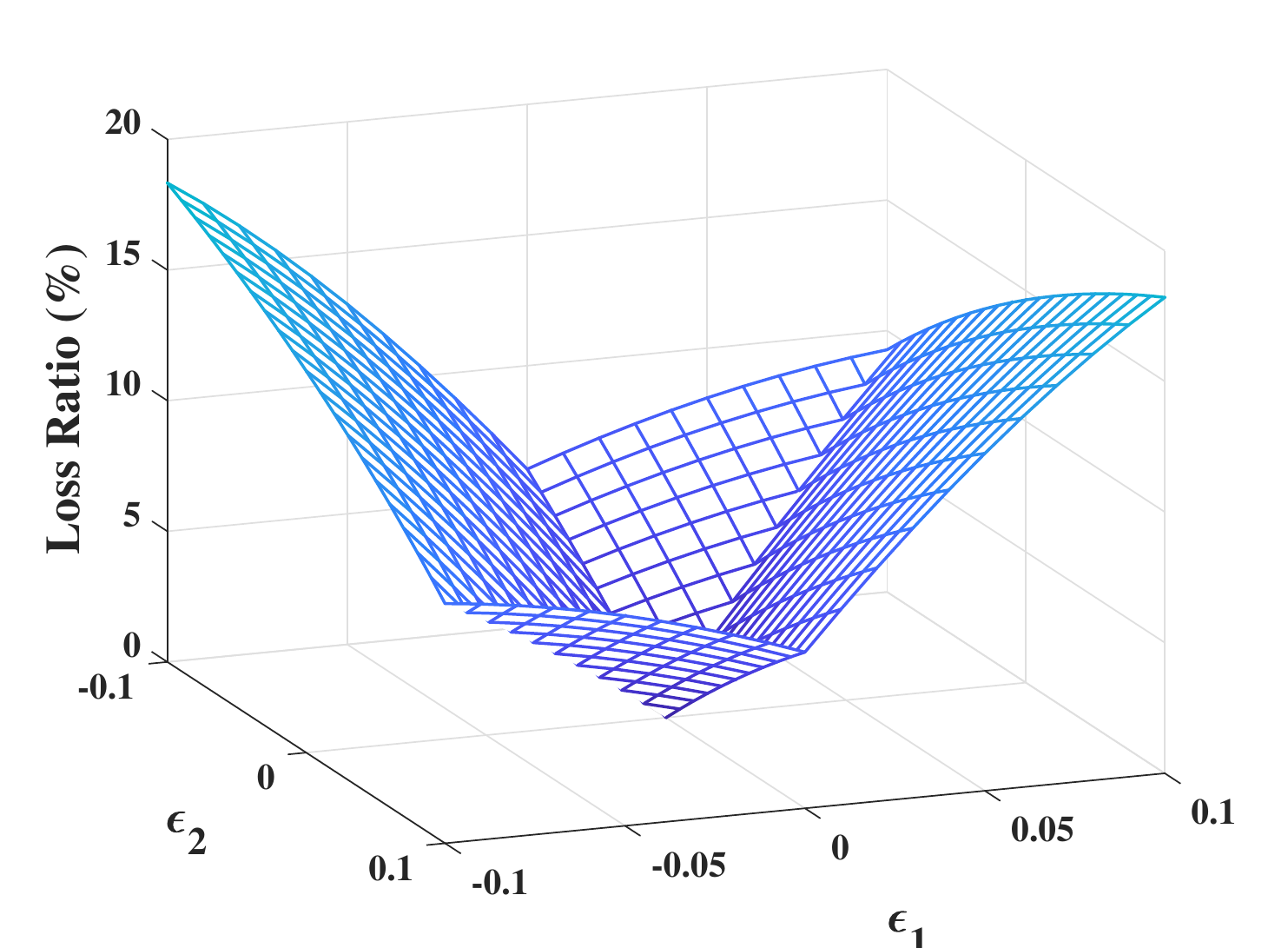}
	\caption{The throughput loss ratio as a function of the normalized synchronization timing error $\epsilon_1$ and the normalized coordination timing error $\epsilon_2$ when $P_1|h_1|^2 = 1$, $P_2|h_2|^2 = 0.5$, $\tau = 0.5$ and $N = 10$.}
	\label{throughput}
\end{figure}
\begin{figure}[t b]
	\centering
	\includegraphics[width=0.9\columnwidth]{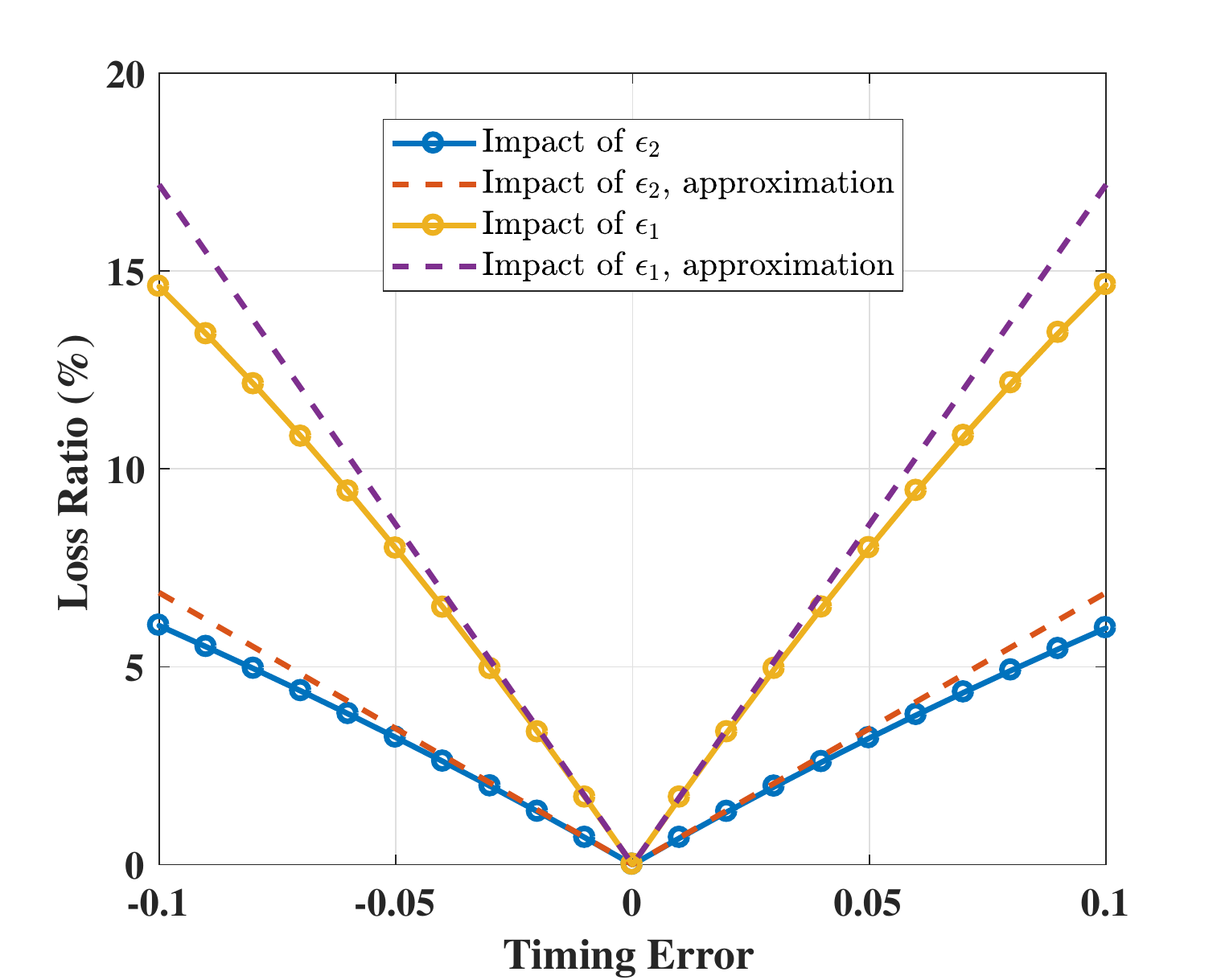}
	\caption{The individual impacts of the normalized synchronization timing error $\epsilon_1$ and the normalized coordination timing error $\epsilon_2$ on the throughput loss ratio when $P_1|h_1|^2 = 1$, $P_2|h_2|^2 = 0.5$, $\tau = 0.5$, and $N = 10$.}
	\label{delta}
\end{figure}
\begin{figure}[t b]
	\centering
	\includegraphics[width=0.9\columnwidth]{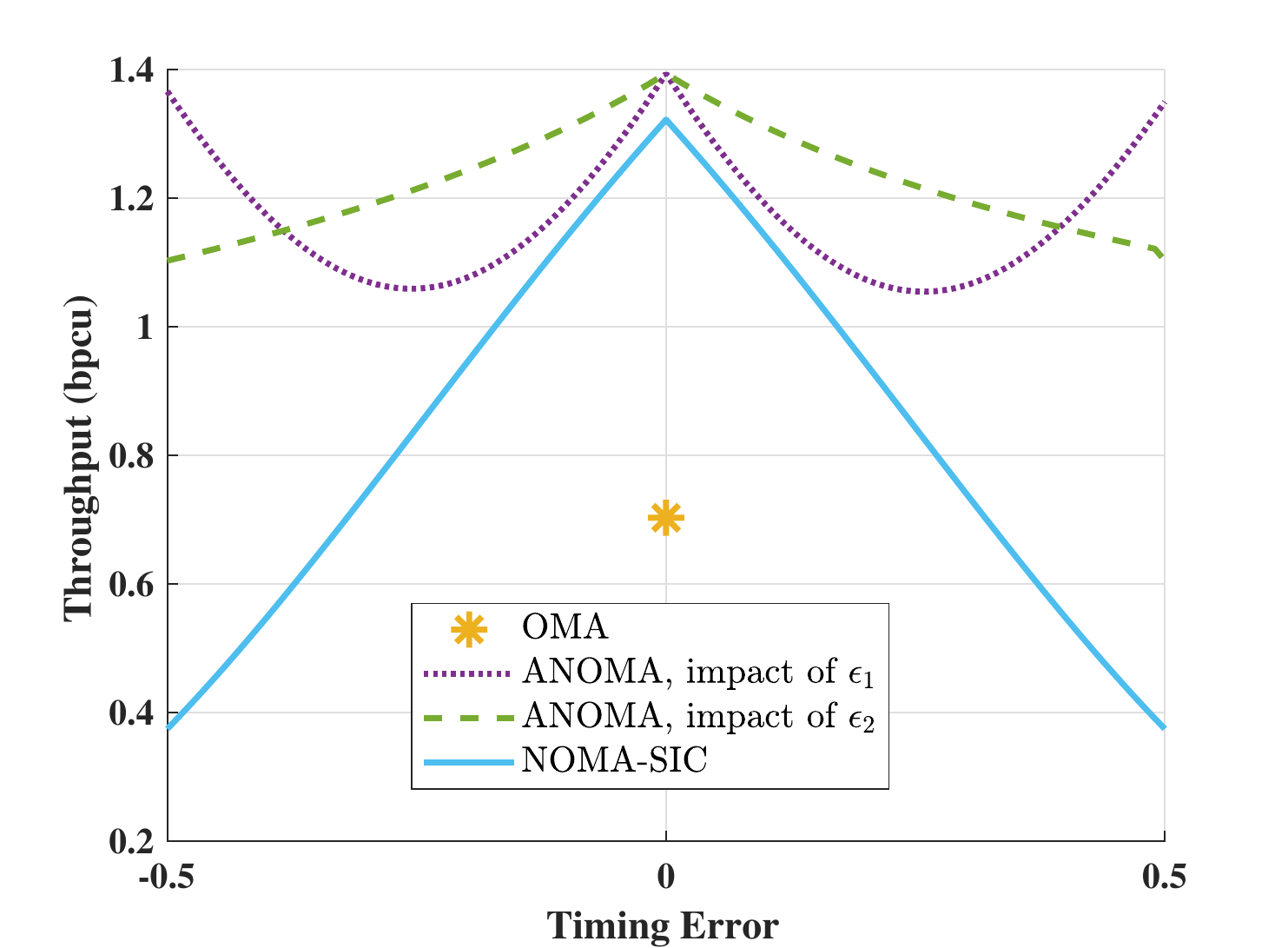}
	\caption{Comparison of throughputs among OMA, NOMA, and ANOMA when $P_1|h_1|^2 = 1$, $P_2|h_2|^2 = 0.5$, $\tau = 0.5$, and $N = 10$.}
	\label{comparison}
\end{figure}

In this section, we present numerical results to compare the throughput performances of NOMA and ANOMA systems and illustrate the impact of timing error on the performance of the ANOMA system. Figs. \ref{compareANOMAandNOMA}, \ref{maxPowerTransmit}, and \ref{optimal_tau} show the ANOMA system without timing error while the other figures are for the impact of timing error. In our simulations, we set the symbol length $T=1$ and the AWGN with unit power. If not specified, the normalized timing mismatch between the two signals $\tau$ is set to 0.5. 

At first, we present the throughput performances of NOMA and ANOMA systems under different channel conditions in Fig.~\ref{rate_vs_h}. In Fig. \ref{rate_vs_h}, the curves of ``ANOMA in \eqref{RsumANOMA}'' are derived directly from the definition in \eqref{RsumANOMA}, and the curves of ``ANOMA in \eqref{eq:theoremRsumANOMA}'' are obtained from our result in Theorem	 \ref{theoremRsumANOMA}. It is shown that the throughput computed by Theorem \ref{theoremRsumANOMA} completely aligns with that calculated by \eqref{RsumANOMA} for different combinations of channel conditions, which confirms the correctness of Theorem \ref{theoremRsumANOMA}. Besides, Fig. \ref{rate_vs_h} demonstrates that the throughputs of ANOMA and NOMA systems increase with the channel gains $|h_1|^2$ and $|h_2|^2$. It is also shown that the throughput of the ANOMA system is better than that of the NOMA system when the frame length $N = 10$ ($N \gg 1$) under different channel conditions.

Then, we compare the throughput performances of NOMA and ANOMA systems. Fig. \ref{compareANOMAandNOMA} shows the throughput as a function of the frame length $N$. In Fig. \ref{compareANOMAandNOMA}, it is demonstrated that as $N$ increases, the throughput of ANOMA systems converges to the result in Corollary \ref{theoremR_NtoInfinity}, which provides the throughput in the asymptotic case of $N \rightarrow \infty$. Furthermore, we note from the figure that the throughputs of ANOMA systems for different $\tau$s as $N \rightarrow \infty$ are greater than that of the NOMA system, which is consistent with our analytical results in Theorem \ref{theoremRanoma>Rnoma} and Corollary \ref{corollarytau*0.5}.

In addition, we illustrate the optimal parameter design of the ANOMA system in Figs. \ref{maxPowerTransmit} and \ref{optimal_tau}. Fig. \ref{maxPowerTransmit} demonstrates the sum throughput of two users as a function of their transmit powers. It is shown that the maximal sum throughput is reached when the transmit powers are equal to the maximum available powers, which aligns with Theorem \ref{FullPowerTheorem}. We present the optimal normalized timing mismatch $\tau^*$ found by exhaustive search to maximize the sum throughput of two users in Fig. \ref{optimal_tau}. As shown in Fig. \ref{optimal_tau}, $\tau^*$ starts with 0, and then increases with $N$, finally converges to 0.5 as $N$ grows, which verifies the correctness of Theorem \ref{theoremtau*0.5}.

In what follows, we evaluate the impact of timing error on the throughput of ANOMA systems. In the following figures, the throughput loss ratio is defined as the ratio of the throughput loss in \eqref{rate_diff} and the throughput of the ANOMA system without timing error in \eqref{RsumANOMA}, i.e.,
\begin{equation}\label{}
\gamma = \frac{\Delta}{R^{\mathrm{ANOMA}}}.
\end{equation}

In Fig. \ref{throughput}, we present the throughput loss ratio as a function of $\epsilon_1$ and $\epsilon_2$ ranging from -0.1 to 0.1. As shown in Fig. \ref{throughput}, the throughput loss ratio increases with both the synchronization timing error and the coordination timing error. We also find that the throughput loss ratio $\gamma$ is a continuous function with respect to $\epsilon_1$ and $\epsilon_2$ but non-differentiable when $\epsilon_1 = 0$ or $\epsilon_1 + \epsilon_2 = 0$. This is because there are non-linear step functions in the expression for $\mathbf{E}_1$ in \eqref{wideeq}.

We also study the individual effects of the timing synchronization error and the coordination timing error on the throughput of ANOMA systems. In Fig. \ref{delta}, we show the throughput loss ratio as a function of $\epsilon_1$ when $\epsilon_2 = 0$ and $\epsilon_2$ when $\epsilon_1 = 0$. Note that the curves of ``impact of $\epsilon_1$'' and ``impact of $\epsilon_2$'' are the slices of Fig. \ref{throughput} when $\epsilon_2 = 0$ and $\epsilon_1 = 0$, respectively.
The approximated results are calculated by $\Delta_{\epsilon_1} / R^{\mathrm{ANOMA}}$ and $\Delta_{\epsilon_2} / R^{\mathrm{ANOMA}}$ using \eqref{delta_e1} and \eqref{delta_e2}. It is demonstrated that the expressions in \eqref{delta_e1} and \eqref{delta_e2} are good approximations of \eqref{rate_diff} when $|\epsilon_1|< 0.05$ and $|\epsilon_2| < 0.05$, respectively. Besides, $\epsilon_1$ causes almost twice throughput loss compared to $\epsilon_2$ for the same value of error. This phenomenon reveals that the performance of ANOMA is more sensitive to the synchronization timing error than the coordination timing error. 

Finally, we compare the performances of OMA, NOMA, ANOMA without and with timing error in Fig. \ref{comparison}. In our simulation, the conventional TDMA is adopted as the OMA scheme. As shown in the figure, the throughput curve of OMA is a single point because it is not a function of timing error. The throughput of the NOMA system is calculated under the assumption that perfect SIC is realized at BS. It is demonstrated that the rate performance for ANOMA without timing error is better than that of NOMA with SIC which is further greater than that of OMA. Also, ANOMA always outperforms a perfectly synchronized OMA. We note that for small values of timing error, ANOMA outperforms even a perfectly synchronized NOMA. For the same timing error, the performance of ANOMA is better than that of NOMA. Besides, as shown in Fig. \ref{comparison}, the throughput decreases with the absolute value of $\epsilon_2$ monotonously, while the throughput decreases at the beginning and then increases as the absolute value of $\epsilon_1$ increases. This phenomenon can be explained as follows: If there is no timing error ($\epsilon_1 = \epsilon_2 = 0$) and $\tau = 0.5$, the sampling moments are at $iT$ and $(i + 0.5)T$, $i = 0, \cdots, N-1$. If $|\epsilon_1| = 0.5$ and $\epsilon_2 = 0$, the sampling moments are at $(i \pm 0.5) T$ and $(i + 1 \pm 0.5)T$, $i = 0, \cdots, N-1$, which are equivalent to advancing ($\epsilon_1 = -0.5$) or delaying ($\epsilon_1 = 0.5$) all sampling moments by $0.5T$. The sampling diversity can still be achieved except that there will be throughput loss due to the shift of sampling moments. For the case $\epsilon_1 = 0$ and $|\epsilon_2| = 0.5$, the second sample vector is a duplicate ($\epsilon_2 = -0.5$) or shifted version ($\epsilon_2 = 0.5$) of the first sample vector. Hence, the sampling diversity cannot be obtained and only the first sample vector can be used to recover the transmitted symbols.

\section{Conclusion}\label{sectionConclusion}
In this paper, we have studied the performance of a two-user uplink ANOMA system and compared it with the NOMA system. We derive an analytical expression for the two-user sum throughput in the ANOMA system as a function of SNR, frame length, and normalized timing mismatch. We have demonstrated that the ANOMA outperforms the NOMA when the frame length is sufficiently large. Furthermore, we have shown that two users should transmit at full power to maximize the two-user sum throughput. The optimal timing mismatch to maximize the sum throughput converges to a half of one time slot as the frame length goes to infinity. Besides, we discuss the impact of timing error on the throughput performance of the ANOMA system. Two types of timing error are taken into consideration, i.e., the synchronization timing error and the coordination timing error. We have shown how these two types of timing error individually and jointly affect the throughput performance of the ANOMA system. We have found that the ANOMA system is more sensitive to the synchronization timing error than the coordination timing error. 

\section*{Acknowledgement}
The authors would like to thank Mehdi Ganji for insightful conversation regarding timing error.

\appendices
\section{Proof of Theorem \ref{theoremRsumANOMA}}\label{ProoftheoremRsumANOMA}
\begin{IEEEproof}
According to \eqref{RsumANOMA}, we can rewrite $\det\left(\mathbf{I}_{2N} + \mathbf{HH}^H\mathbf{R}\right)$ as
\begin{align}\label{det_separate}
\det\left(\mathbf{I}_{2N} + \mathbf{HH}^H\mathbf{R}\right) &= \det\left(\mathbf{HH}^H\right)\det\left((\mathbf{HH}^H)^{-1} + \mathbf{R}\right)\notag\\
&= \left(P_1|h_1|^2\right)^N\left(P_2|h_2|^2\right)^N\det\left((\mathbf{HH}^H)^{-1} + \mathbf{R}\right).
\end{align}

According to \eqref{R_matrix} and \eqref{H_matrix}, $(\mathbf{HH}^H)^{-1} + \mathbf{R}$ is a $2N \times 2N$ matrix calculated by
\begin{align}
(\mathbf{HH}^H)^{-1} + \mathbf{R} = \left[\begin{smallmatrix}
1 + \left(P_1|h_1|^2\right)^{-1}\ &1-\tau\ &0\ &\cdots\ &\cdots\ &0\\
1-\tau\ &1+\left(P_2|h_2|^2\right)^{-1}\ &\tau\ &0\ &\cdots\ &0\\
\vdots\ &\ddots\ &\ddots\ &\ddots\ &\ddots\ &\vdots\\
0\ &\cdots\ &1-\tau\ &1+\left(P_2|h_2|^2\right)^{-1}\ &\tau\ &0\\
0\ &\cdots\ &0\ &\tau\ &1+\left(P_1|h_1|^2\right)^{-1}\ &1-\tau\\
0\ &\cdots\ &\cdots\ &0\ &1-\tau\ &1+\left(P_2|h_2|^2\right)^{-1}
\end{smallmatrix}\right].
\end{align}

For simplicity of presentation, we define $\mu_1 = P_1|h_1|^2$, $\mu_2 = P_2|h_2|^2$, and
\begin{align}
d_m =
\left\{
\begin{array}{lr}
\det \left(\left[\begin{smallmatrix}
1 + \mu_1^{-1}\ &1-\tau\ &0\ &\cdots\ &\cdots\ &0\\
1-\tau\ &1+\mu_2^{-1}\ &\tau\ &0\ &\cdots\ &0\\
\vdots\ &\ddots\ &\ddots\ &\ddots\ &\ddots\ &\vdots\\
0\ &\cdots\ &1-\tau\ &1+\mu_2^{-1}\ &\tau\ &0\\
0\ &\cdots\ &0\ &\tau\ &1+\mu_1^{-1}\ &1-\tau\\
0\ &\cdots\ &\cdots\ &0\ &1-\tau\ &1+\mu_2^{-1}
\end{smallmatrix}\right]_{m\times m}\right), \ & \mathrm{if}\ m\mathrm{\ is\ even},\\
\det\left(\left[\begin{smallmatrix}
1 + \mu_1^{-1}\ &0\ &1-\tau\ &\cdots\ &\cdots\ &0\\
1-\tau\ &1+\mu_2^{-1}\ &\tau\ &0\ &\cdots\ &0\\
\vdots\ &\ddots\ &\ddots\ &\ddots\ &\ddots\ &\vdots\\
0\ &\cdots\ &\tau\ &1+\mu_1^{-1}\ &1-\tau\ &0\\
0\ &\cdots\ &0\ &1-\tau\ &1+\mu_2^{-1}\ &\tau\\
0\ &\cdots\ &\cdots\ &0\ &\tau\ &1+\mu_1^{-1}\\
\end{smallmatrix}\right]_{m\times m}\right), & \mathrm{if}\  m\mathrm{\ is\ odd}.
\end{array}
\right.
\end{align}

Thus, 
\begin{align}
\det\left((\mathbf{HH}^H)^{-1} + \mathbf{R}\right)=d_{2N}.
\end{align}

By the method of cofactor expansion \cite{poole2014linear}, the determinant of $\det\left((\mathbf{HH}^H)^{-1} + \mathbf{R}\right)$ can be expressed as a weighted sum of the determinants of its minors. The minor $M_{i, j}$ is defined as the determinant of the matrix that results from $(\mathbf{HH}^H)^{-1} + \mathbf{R}$ by removing the $i$th row and the $j$th column. Then, we have  
\begin{align}\label{d2Nrecursion}
d_{2N}
&= \sum_{j = 1}^{2N} (-1)^{2N + j} a_{2N, j} M_{2N, j}\notag\\
&= (-1)^{2N+2N}\left(1+\mu_2^{-1}\right) \underbrace{\det\left(\left[\begin{smallmatrix}
1 + \mu_1^{-1}\ &1-\tau\ &\cdots\ &\cdots\ &0\\
1-\tau\ &1+\mu_2^{-1}\ &\tau\ &\cdots\ &0\\
\vdots\ &\ddots\ &\ddots\ &\ddots\ &\vdots\\
0\ &\cdots\ &1-\tau\ &1+\mu_2^{-1}\ &\tau\\
0\ &\cdots\ &0\ &\tau\ &1+\mu_1^{-1}\\
\end{smallmatrix}\right]_{(2N-1)\times(2N-1)}\right)}_{d_{2N-1}}\notag\\
&+ (-1)^{2N+2N-1}(1-\tau)\det\left(\left[\begin{smallmatrix}
1 + \mu_1^{-1}\ &1-\tau\ &\cdots\ &\cdots\ &0\\
1-\tau\ &1+\mu_2^{-1}\ &\tau\ &\cdots\ &0\\
\vdots\ &\ddots\ &\ddots\ &\ddots\ &\vdots\\
0\ &\cdots\ &1-\tau\ &1+\mu_2^{-1}\ &0\\
0\ &\cdots\ &0\ &\tau\ &1-\tau\\
\end{smallmatrix}\right]_{(2N-1)\times(2N-1)}\right)\notag\\
&= \left(1+\mu_2^{-1}\right)d_{2N-1}\notag\\
&- (1-\tau)^2(-1)^{4N-2}\underbrace{\det\left(\left[\begin{smallmatrix}
1 + \mu_1^{-1}\ &1-\tau\ &0\ &\cdots\ &\cdots\ &0\\
1-\tau\ &1+\mu_2^{-1}\ &\tau\ &0\ &\cdots\ &0\\
\vdots\ &\ddots\ &\ddots\ &\ddots\ &\ddots\ &\vdots\\
0\ &\cdots\ &1-\tau\ &1+\mu_2^{-1}\ &\tau\ &0\\
0\ &\cdots\ &0\ &\tau\ &1+\mu_1^{-1}\ &1-\tau\\
0\ &\cdots\ &\cdots\ &0\ &1-\tau\ &1+\mu_2^{-1}
\end{smallmatrix}\right]_{(2N-2)\times(2N-2)}\right)}_{d_{2N-2}}\notag\\
&= \left(1+\mu_2^{-1}\right)d_{2N-1} - (1-\tau)^2d_{2N-2},
\end{align}
where $N \ge 2$ and $a_{i, j}$ denotes the element of the matrix $(\mathbf{HH}^H)^{-1} + \mathbf{R}$ at the $i$th row and the $j$th column.
Similarly, we can also write the recursive formula for $d_{2N-1}$ as
\begin{align}\label{d2N-1recursion}
d_{2N-1} = (1 + \mu_1^{-1}) d_{2N-2} - \tau^2 d_{2N-3}.
\end{align}

From \eqref{d2Nrecursion} and \eqref{d2N-1recursion}, we obtain
\begin{align}\label{d2Nrecursion2}
d_{2N} = \left[\mu_1^{-1} + \mu_2^{-1} + \mu_1^{-1}\mu_2^{-1} + 2\tau(1 - \tau)\right] d_{2N-2} - \tau^2(1 - \tau)^2d_{2N-4}.
\end{align}

To formalize  \eqref{d2Nrecursion2} as the recursion formula of a geometric progression,  \eqref{d2Nrecursion2} can be rewritten as
\begin{align}
d_{2N} - r_1d_{2N-2} &= r_2(d_{2N-2} - r_1d_{2N-4}),\label{recursiongroup1}\\
d_{2N} - r_2d_{2N-2} &= r_1(d_{2N-2} - r_2d_{2N-4}),\label{recursiongroup2}
\end{align}
where
\begin{equation}\label{eq:r1inproof}
r_1 =
	\frac{\mu_1^{-1}\! +\! \mu_2^{-1}\! +\! \mu_1^{-1}\mu_2^{-1}\! +\! 2\tau(1-\tau)\! +\! \sqrt{\left[\mu_1^{-1} + \mu_2^{-1} + \mu_1^{-1}\mu_2^{-1} + 2\tau(1-\tau)\right]^2\! -\! 4\tau^2(1-\tau)^2}}{2},
\end{equation}
\begin{equation}\label{eq:r2inproof}
r_2 =
\frac{\mu_1^{-1}\! +\! \mu_2^{-1}\! +\! \mu_1^{-1}\mu_2^{-1}\! +\! 2\tau(1-\tau)\! -\! \sqrt{\left[\mu_1^{-1} + \mu_2^{-1} + \mu_1^{-1}\mu_2^{-1} + 2\tau(1-\tau)\right]^2\! -\! 4\tau^2(1-\tau)^2}}{2}.
\end{equation}

Since $\mu_1 > 0$, $\mu_2 > 0$, and $\tau\in [0, 1)$, we note that the part under the square root symbol in \eqref{eq:r1inproof} and \eqref{eq:r2inproof} is always positive, such that
\begin{align}\label{sqrtterm}
&\left[\mu_1^{-1} + \mu_2^{-1} + \mu_1^{-1}\mu_2^{-1} + 2\tau(1-\tau)\right]^2 - 4\tau^2(1-\tau)^2\notag\\
&= \left[\mu_1^{-1} + \mu_2^{-1} + \mu_1^{-1}\mu_2^{-1} + 4\tau(1-\tau)\right]\left[\mu_1^{-1} + \mu_2^{-1} + \mu_1^{-1}\mu_2^{-1}\right] > 0.
\end{align}

From \eqref{recursiongroup1} and \eqref{recursiongroup2}, we obtain
\begin{align}
d_{2N} - r_1d_{2N-2} &= r_2^{N-1}(d_{2} - r_1d_{0}),\label{eqgroup1}
\end{align}
and
\begin{align}
d_{2N} - r_2d_{2N-2} &= r_1^{N-1}(d_{2} - r_2d_{0}).\label{eqgroup2}
\end{align}

Solving $d_{2N}$ from the equation group constituted by  \eqref{eqgroup1} and \eqref{eqgroup2}, we derive
\begin{align}\label{d2Nfinalrecursion}
d_{2N} = \frac{r_1^N(d_2 - r_2d_0) - r_2^N(d_2 - r_1d_0)}{r_1 - r_2}.
\end{align}

Substituting $d_0 = 1$ and
\begin{align}
d_2 = \left[\begin{matrix}
1 + \mu_1^{-1}\ &1-\tau\\
1-\tau\ &1 + \mu_2^{-1}
\end{matrix}\right] = \left(1 + \mu_1^{-1}\right)\left(1 + \mu_2^{-1}\right) - (1-\tau)^2 = r_1 + r_2 + \tau^2
\end{align}
 into \eqref{d2Nfinalrecursion}, we have
\begin{align}\label{eq:d2nfinalex}
d_{2N} = \frac{(r_1^{N+1} - r_2^{N+1}) + \tau^2(r_1^N - r_2^N)}{r_1 - r_2}.
\end{align}

Finally, based on \eqref{det_separate} and \eqref{eq:d2nfinalex}, we obtain the throughput as
\begin{align}
R^{\mathrm{ANOMA}} = \frac{N}{N+\tau}\log\left(\mu_1\mu_2\right) + \frac{1}{N+\tau}\log \frac{(r_1^{N+1} - r_2^{N+1}) + \tau^2(r_1^N - r_2^N)}{r_1 - r_2}.
\end{align}

This completes the proof.
\end{IEEEproof}

\section{Proof of Corollary \ref{theoremR_NtoInfinity}}\label{ProoftheoremR_NtoInfinity}
\begin{IEEEproof}
	Note that $\mu_1$, $\mu_2$, $r_1$, $r_2$, and $\tau$ are all independent of $N$. We then have
	\begin{align}
		\lim_{N\rightarrow \infty} R^{\mathrm{ANOMA}} &= \lim_{N\rightarrow \infty} \frac{N}{N+\tau}\log (\mu_1\mu_2) + \frac{\log \left[(r_1^{N+1} - r_2^{N+1}) + \tau^2(r_1^N - r_2^N)\right] - \log (r_1 - r_2)}{N+\tau}\notag\\
		&\stackrel{(a)}{=} \log (\mu_1\mu_2) + \lim_{N\rightarrow \infty} \frac{(r_1^{N+1}\log r_1 - r_2^{N+1}\log r_2) + \tau^2(r_1^N\log r_1 - r_2^N \log r_2)}{(r_1^{N+1} - r_2^{N+1}) + \tau^2(r_1^N - r_2^N)}\notag\\
		&\stackrel{(b)}{=} \log (\mu_1\mu_2) + \lim_{N\rightarrow \infty} \frac{(r_1\alpha^N\log r_1 - r_2\log r_2) + \tau^2(\alpha^N\log r_1 - \log r_2)}{(r_1\alpha^N - r_2) + \tau^2(\alpha^N - 1)}\notag\\
		&\stackrel{}{=} \log (\mu_1\mu_2) + \lim_{N\rightarrow \infty} \frac{\alpha^N(r_1 + \tau^2)\log r_1 - (r_2 + \tau^2)\log r_2}{\alpha^N(r_1 + \tau^2) - (r_2 + \tau^2)}\notag\\
		& \stackrel{(c)}{=} \log \left(\mu_1\mu_2r_1\right),
	\end{align}
	where $\alpha=r_1/r_2$, $(a)$ is derived by applying L'Hospital's rule,	$(b)$ is derived by dividing both the numerator and the denominator by $r_2^N$, and $(c)$ is obtained from the facts that $r_1 > r_2 > 0$ and $\alpha > 1$ according to \eqref{r_1} and \eqref{r_2}.
	This completes the proof.
\end{IEEEproof}

\section{Proof of Theorem \ref{theoremRanoma>Rnoma}}\label{ProofRanoma>Rnoma}
\begin{IEEEproof}
	The expressions for $\lim_{N\rightarrow \infty} R^{\mathrm{ANOMA}}$ and $R^{\mathrm{NOMA}}$ are given by
	\begin{align}\label{proofExpressionRanoma}
	\lim_{N\rightarrow \infty} R^{\mathrm{ANOMA}} &= \log \left(\mu_1\mu_2 r_1\right)\notag\\
	&= \log \left\{\frac{1 + \mu_1 + \mu_2 + \mu_1\mu_2(2\tau - 2\tau^2)}{2}\right.\notag\\
	& + \left.\frac{\sqrt{\left(1+\mu_1+\mu_2\right)^2 + 2 \left(1+\mu_1 +\mu_2\right)\mu_1\mu_2(2\tau - 2\tau^2)}}{2}\right\}
	\end{align}
	and $R^{\mathrm{NOMA}} = \log (1 + \mu_1 + \mu_2)$, respectively. 
	
	If $\tau = 0$, it is easy to find that
	\begin{align}
	\lim_{N\rightarrow \infty} R^{\mathrm{ANOMA}} = \log (1 + \mu_1 + \mu_2) = R^{\mathrm{NOMA}}.
	\end{align}
	
	If $\tau \ne 0$, i.e., $\tau \in (0, 1)$, we have $2\tau - 2\tau^2 > 0$. According to \eqref{proofExpressionRanoma}, since $\mu_1 > 0$ and $\mu_2 > 0$, we obtain
	\begin{align}
	\lim_{N\rightarrow \infty} R^{\mathrm{ANOMA}} &= \log \left\{\frac{1 + \mu_1 + \mu_2 + \mu_1\mu_2(2\tau - 2\tau^2)}{2}\right.\notag\\
	& + \left.\frac{\sqrt{\left(1+\mu_1+\mu_2\right)^2 + 2 \left(1+\mu_1 +\mu_2\right)\mu_1\mu_2(2\tau - 2\tau^2)}}{2}\right\}\notag\\
	&> \log \left\{\frac{1 + \mu_1 + \mu_2}{2} + \frac{\sqrt{\left(1+\mu_1+\mu_2\right)^2 }}{2}\right\} = R^{\mathrm{NOMA}}.
	\end{align}

Until now, we have proved $\lim_{N\rightarrow \infty} R^{\mathrm{ANOMA}} = R^{\mathrm{NOMA}}$ if $\tau = 0$ and $\lim_{N\rightarrow \infty} R^{\mathrm{ANOMA}} > R^{\mathrm{NOMA}}$ if $\tau \neq 0$. Next, we need to prove $\tau = 0$ if $\lim_{N\rightarrow \infty} R^{\mathrm{ANOMA}} = R^{\mathrm{NOMA}}$.

If $\lim_{N\rightarrow \infty}R^{\mathrm{ANOMA}} = R^{\mathrm{NOMA}}$, we have
\begin{align}
\lim_{N\rightarrow \infty}R^{\mathrm{ANOMA}} = \log(\mu_1\mu_2 r_1) &= \log(1 + \mu_1 + \mu_2) = R^{\mathrm{NOMA}}.
\end{align}

After simplifications, we have
\begin{align}\label{eq:Ranoma=Rnoma}
\sqrt{\left(1\!+\!\mu_1\!+\!\mu_2\right)^2\! +\! 2 \left(1\!+\!\mu_1\! +\!\mu_2\right)\mu_1\mu_2(2\tau - 2\tau^2)} &= 1\! +\! \mu_1\! +\! \mu_2\! -\!\mu_1\mu_2(2\tau - 2\tau^2).
\end{align}

Note that \eqref{eq:Ranoma=Rnoma} holds only if the right side of \eqref{eq:Ranoma=Rnoma} is non-negative, i.e., 
\begin{align}\label{eq:Ranoma=RnomaPrecondition}
1 + \mu_1 + \mu_2 -\mu_1\mu_2(2\tau - 2\tau^2) \ge 0.
\end{align}

Squaring both sides of the equal sign in \eqref{eq:Ranoma=Rnoma}, we obtain
\begin{align}\label{eq:Ranoma=Rnoma2}
4(1 + \mu_1 + \mu_2)(2\tau - 2\tau^2) = \mu_1\mu_2(2\tau - 2\tau^2)^2.
\end{align}

Then, \eqref{eq:Ranoma=Rnoma2} holds if $2\tau - 2\tau^2 = 0$ or $4(1 + \mu_1 + \mu_2) = \mu_1\mu_2(2\tau - 2\tau^2)$. It is easy to prove that $4(1 + \mu_1 + \mu_2) = \mu_1\mu_2(2\tau - 2\tau^2)$ contradicts \eqref{eq:Ranoma=RnomaPrecondition}. As a result, \eqref{eq:Ranoma=Rnoma2} holds only if $2\tau - 2\tau^2 = 0$ which then leads to $\tau = 0$. 

	Therefore, $\lim_{N\rightarrow \infty}R^{\mathrm{ANOMA}} \ge R^{\mathrm{NOMA}}$ is always true and the equal sign is achieved if and only if $\tau = 0$. This completes the proof.
\end{IEEEproof}

\section{Proof of Theorem \ref{FullPowerTheorem}}\label{ProofFullPowerTheorem}
\begin{IEEEproof}
	From Theorem~\ref{theoremRsumANOMA}, we have
	\begin{align}\label{Proof2Sumrate}
	R^{\mathrm{ANOMA}} &= \frac{N}{N+\tau}\log\left(\mu_1\mu_2\right) + \frac{1}{N+\tau}\log \frac{(r_1^{N+1} - r_2^{N+1}) + \tau^2(r_1^N - r_2^N)}{r_1 - r_2}\notag\\
	&\stackrel{(a)}{=} \frac{N}{N+\tau}\log\left(\mu_1\mu_2\right) + \frac{1}{N+\tau}\log \left[\sum_{i=0}^{N}r_1^i r_2^{N-i} + \tau^2\sum_{i=0}^{N-1}r_1^ir_2^{N-1-i}\right]\notag\\
	&= \frac{1}{N+\tau}\log \left[\sum_{i=0}^{N}\mu_1^N \mu_2^N r_1^i r_2^{N-i} + \tau^2\sum_{i=0}^{N-1} \mu_1^N \mu_2^N r_1^ir_2^{N-1-i}\right]\notag\\
	&= \frac{1}{N+\tau}\log \left[\sum_{i=0}^{N}(\mu_1\mu_2)^{N-i} (\mu_1\mu_2r_1)^i r_2^{N-i} + \tau^2\sum_{i=0}^{N-1} (\mu_1\mu_2)^{N-i} (\mu_1\mu_2r_1)^ir_2^{N-1-i}\right],
	\end{align}
	where $(a)$ is derived by applying $a^N - b^N = (a - b)(\sum_{i=0}^{N-1}a^ib^{N-1-i})$. In what follows, we prove that $r_2$ is a non-decreasing function of  $\mu_1$
and $\mu_2$, and $\mu_1\mu_2r_1$ increases as $\mu_1$
and $\mu_2$ increase, so that  $R^{\mathrm{ANOMA}}$  increases as $\mu_1$
and $\mu_2$ increase.

From \eqref{r_1}, we can find that
	\begin{align}
	\frac{\partial r_1}{\partial \mu_1} < 0\ \mathrm{and}\ \frac{\partial r_1}{\partial \mu_2} < 0.
	\end{align}

Since $r_2 = \tau^2(1-\tau)^2/r_1$, we further find that
	\begin{align}\label{Proof2derivativer2}
	\frac{\partial r_2}{\partial \mu_1} = -\frac{\tau^2(1-\tau)^2}{r_1^2}\frac{\partial r_1}{\partial \mu_1} \geqslant 0\ \mathrm{and}\ \frac{\partial r_2}{\partial \mu_2} = -\frac{\tau^2(1-\tau)^2}{r_1^2}\frac{\partial r_1}{\partial \mu_2} \geqslant 0.
	\end{align}

With \eqref{r_1}, we have
	\begin{align}\label{Proof2mu_1mu_2r_1}
	\mu_1\mu_2 r_1 &= \frac{1 + \mu_1 + \mu_2 + \mu_1\mu_2(2\tau - 2\tau^2)}{2}\notag\\
	& + \frac{\sqrt{\left(1+\mu_1+\mu_2\right)^2 + 2 \left(1+\mu_1 +\mu_2\right)\mu_1\mu_2(2\tau - 2\tau^2)}}{2}.
	\end{align}

From  \eqref{Proof2mu_1mu_2r_1}, we can derive that
	\begin{align}\label{Proof2derivative_mu_1mu_2r_1}
	\frac{\partial (\mu_1\mu_2r_1)}{\partial \mu_1} > 0\ \mathrm{and}\ \frac{\partial (\mu_1\mu_2r_1)}{\partial \mu_2} > 0.
	\end{align}

Based on \eqref{Proof2derivativer2} and \eqref{Proof2derivative_mu_1mu_2r_1}, we note that $r_2$ is a non-decreasing function of  $\mu_1$ and $\mu_2$, and $\mu_1\mu_2r_1$ increases as $\mu_1$ and $\mu_2$ increase. In addition, since $\mu_1$, $\mu_2$, $r_2$, and $\mu_1\mu_2 r_1$ are positive, the term $(\mu_1\mu_2)^{j}(\mu_1\mu_2 r_1)^i r_2^{M-i}$ ($i=0,\cdots,M-1$, $j \ge 0$) is an increasing function of $\mu_1$ and $\mu_2$ for any positive $M$. Then, $R^{\mathrm{ANOMA}}$ is an increasing function of $\mu_1$ and $\mu_2$ because it is constituted by a sum of $(\mu_1\mu_2)^{j}(\mu_1\mu_2 r_1)^i r_2^{M-i}$ ($i=0,\cdots,M-1$, $j \ge 0$, $M > 0$). Hence, maximizing the throughput is equivalent to maximizing $\mu_1$ and $\mu_2$ simultaneously, which means that the two users should transmit at full power. This completes the proof.
\end{IEEEproof}

\section{Proof of Theorem \ref{theoremtau*0.5}}\label{prooftheoremtau*0.5}
\begin{IEEEproof}
	\begin{align}
	\tau^* &= \mathop{\arg\max}_{\tau}\lim_{N\rightarrow \infty} R^{\mathrm{ANOMA}}\notag\\
	&= \mathop{\arg\max}_{\tau} \log\left(\mu_1\mu_2 r_1\right)\notag\\
	&= \mathop{\arg\max}_{\tau} \log \left\{\frac{1 + \mu_1 + \mu_2 + \mu_1\mu_2(2\tau - 2\tau^2)}{2}\right.\notag\\
	& + \left.\frac{\sqrt{\left(1+\mu_1+\mu_2\right)^2 + 2 \left(1+\mu_1 +\mu_2\right)\mu_1\mu_2(2\tau - 2\tau^2)}}{2}\right\}\notag\\
	&\stackrel{(a)}{=} \mathop{\arg\max}_{\tau} \left[2\tau - 2\tau^2\right]\notag\\
	&= 0.5,
	\end{align}
	where $(a)$ is derived due to the fact that $\mu_1$ and $\mu_2$ are positive and independent of $\tau$. This completes the proof.
\end{IEEEproof}

\end{document}